\documentclass[twocolumn,pre,superscriptaddress,floatfix,amsmath,amssymb,aps]{revtex4-2}
\pdfoutput=1

\usepackage[utf8]{inputenc}
\usepackage{natbib}
\usepackage{array}
\usepackage{booktabs}
\usepackage{graphicx}
\usepackage{paralist}
\usepackage{color,soul}
\usepackage[colorlinks=true, allcolors=blue]{hyperref}
\usepackage{mathtools}
\usepackage{amsmath}
\usepackage{dcolumn}
\usepackage{amssymb}
\usepackage{gensymb}
\usepackage{mathdesign}
\usepackage{siunitx}
\usepackage{accents}
\usepackage{bm}
\usepackage{dcolumn}

\begin{document}

\title{Large-scale frictionless jamming with power-law particle size distributions}

\author{Joseph M. Monti}
\affiliation{Sandia National Laboratories, Albuquerque, NM 87185, USA}
\author{Joel T. Clemmer}
\affiliation{Sandia National Laboratories, Albuquerque, NM 87185, USA}
\author{Ishan Srivastava}
\affiliation{Center for Computational Sciences and Engineering, Lawrence Berkeley National Laboratory, Berkeley, California 94720, USA}
\author{Leonardo E. Silbert}
\affiliation{School of Math, Science, and Engineering, Central New Mexico Community College, Albuquerque, New Mexico 87106, USA}
\author{Gary S. Grest}
\affiliation{Sandia National Laboratories, Albuquerque, NM 87185, USA}
\author{Jeremy B. Lechman}
\affiliation{Sandia National Laboratories, Albuquerque, NM 87185, USA}
\date{\today}

\begin{abstract}
Due to significant computational expense, discrete element method simulations of jammed packings of size-dispersed spheres with size ratios greater than 1:10 have remained elusive, limiting the correspondence between simulations and real-world granular materials with large size dispersity. Invoking a recently developed neighbor binning algorithm, we generate mechanically-stable jammed packings of frictionless spheres with power-law size distributions containing up to nearly four million particles with size ratios up to 1:100. By systematically varying the width and exponent of the underlying power laws, we analyze the role of particle size distributions on the structure of jammed packings. The densest packings are obtained for size distributions that balance the relative abundance of large-large/intermediate and small-small particle contacts. Although the proportion of rattler particles and mean coordination number strongly depend on the size distribution, the mean coordination of non-rattler particles attains the frictionless isostatic value of six in all cases. The size distribution of non-rattler particles that participate in the load-bearing network exhibits no dependence on the width of the total particle size distribution beyond a critical particle size for low-magnitude exponent power laws. This signifies that only particles with sizes greater than the critical particle size contribute to the mechanical stability. However, for high-magnitude exponent power laws, all particle sizes participate in the mechanical stability of the packing.
\end{abstract}

\maketitle

\section{Introduction}

Packings of stiff granular particles with a high degree of size dispersity are of widespread geophysical and industrial relevance, with applications including powder technology and the mechanics of soil and construction materials~\cite{fuller1907,andreasen1930,furnas1931,turcotte1986,langston1997,liu2019}.
The distribution of particle sizes can adopt discrete or continuous forms, both of which have been shown for frictionless particles to produce overall packing densities, $\phi$, that are greater than the frictionless, monodisperse value $\phi^{\rm mono}\approx 0.64$~\cite{ohern2003}.
The simplest discrete form is the bidisperse case, for which Furnas~\cite{furnas1931} predicted the theoretical limiting value of $\phi^{\rm bi}\approx 0.87$ for an infinitely large size ratio; recent large-scale numerical simulations of bidisperse packings produced packing densities approaching the Furnas limit~\cite{farr2009,srivastava2021}.
To date, most three-dimensional (3D) numerical simulations of continuous, highly disperse systems with size distributions of diverse functional forms have been limited to largest-to-smallest particle size ratios of order 10 or less, and typically only reach packing densities $\sim\!0.71$ or smaller at low confining pressures~\cite{farr2009,danisch2010,hermes2010,desmond2014,baranau2014,cantor2018,mutabaruka2019}. 
To our knowledge, the main exception is the work of~\citet{oquendo2020,oquendo2021,oquendo2022}, who considered power-law-\textit{like} particle size distributions.
These distributions were generated by matching the scaling behavior of the cumulative particle size distributions to early experimental observations by~\citet{fuller1907}.
~\citet{oquendo2020,oquendo2021,oquendo2022} simulated particle size ratios of up to 32, and achieved packing densities close to $0.86$ depending on the characteristics of the particle size distribution.

Here, we consider power-law particle size distributions to study packings of highly disperse particles.
From a numerical perspective, power laws are one of the simplest continuous distributions, since there are only two parameters governing the distribution: the maximum particle size ratio and the power-law exponent.
Power-law distributions have been measured to emerge naturally from various fragmentation mechanisms, including that of sea-ice floes~\cite{herman2013} and comminution~\cite{turcotte1986,filgueira2006,ben-nun2010,minh2013,debono2020}.
Further, power laws display scale invariance and fractal behavior; the geometric Apollonian packing is one such example of a fractal packing with an underlying power-law size distribution~\cite{borkovec1994,anishchik1995,herrmann2003,varrato2011}.

One of the challenges associated with simulating power-law distributions is that the tail of the distribution has significant weight and decays slowly, thus requiring simulating very large size ratios of particles to accurately sample the distribution.
Computational costs of 3D discrete element method (DEM) simulations with broad particle size disparities have been prohibitive until recently due to algorithmic limitations.
Conventional neighbor list generation methods, e.g., those available by default in popular molecular dynamics (MD) packages like LAMMPS~\cite{plimpton1995,thompson2022} exhibit poor scaling with increasing size ratios and become intractable beyond particle size dispersity of order $10$~\cite{veld2008}.
~\citet{ogarko2012} recently developed an improved neighbor list generation scheme. 
A similar approach has since been implemented in LAMMPS by~\citet{shire2021}.
This implementation has been expanded upon and was used to study bidisperse packings of both frictionless and frictional particles with particle size ratios of up to 40~\cite{srivastava2021}.
For our study, we have exerted this simulation capability further to investigate strongly disperse power-law-distributed systems of frictionless particles with unprecedented particle size ratios of up to 100.

Packings of highly disperse particles require careful treatment as relaxation may occur over disparate time scales. 
Unlike volume-controlled jamming protocols, pressure-controlled jamming protocols are guaranteed to produce mechanically-stable packings, and yield greater accessibility to the jamming point in the low pressure regime~\cite{dagois2012,smith2014}.
Recently, the isobaric-isoenthalpic (NPH) thermodynamic ensemble, one example of a constant-pressure protocol, was successfully applied to multi-friction-mode monodisperse packings~\cite{santos2020} and to frictionless and frictional bidisperse packings~\cite{srivastava2021}.
The NPH ensemble implementation in LAMMPS can be leveraged to enforce the condition of zero shear stresses applied to the simulation box concurrently with isotropic compression.
For the highly disperse systems considered in this work, permitting the relaxation to zero of the off-diagonal components of the internal stress tensor, $\mathbf{P}_{\rm int}$, is crucial as this technique produces packings that are also stable with respect to shear deformations.

This article describes numerical simulations performed using a constant-pressure (NPH) compression protocol to generate jammed packings of frictionless, power-law-distributed disperse spherical particles.
The interparticle contact model is described in Section~\ref{sec:cmodel}.
Section~\ref{sec:distro} elucidates how distributions of particle sizes are generated and characterized.
Section~\ref{sec:multi} contains a brief description of the multi-neighboring scheme used in this work.
Further details and benchmark results can be found in the Appendix.
The packing protocol is described in Section~\ref{sec:protocol}.
In Section~\ref{sec:charpack}, the results of packed systems of power-law-distributed particles are characterized.
Finally, Section~\ref{sec:norattlers} examines several properties of the resultant force-bearing networks.

\section{Methods}
%
\begin{figure*}
\centering
\includegraphics[width=1\textwidth]{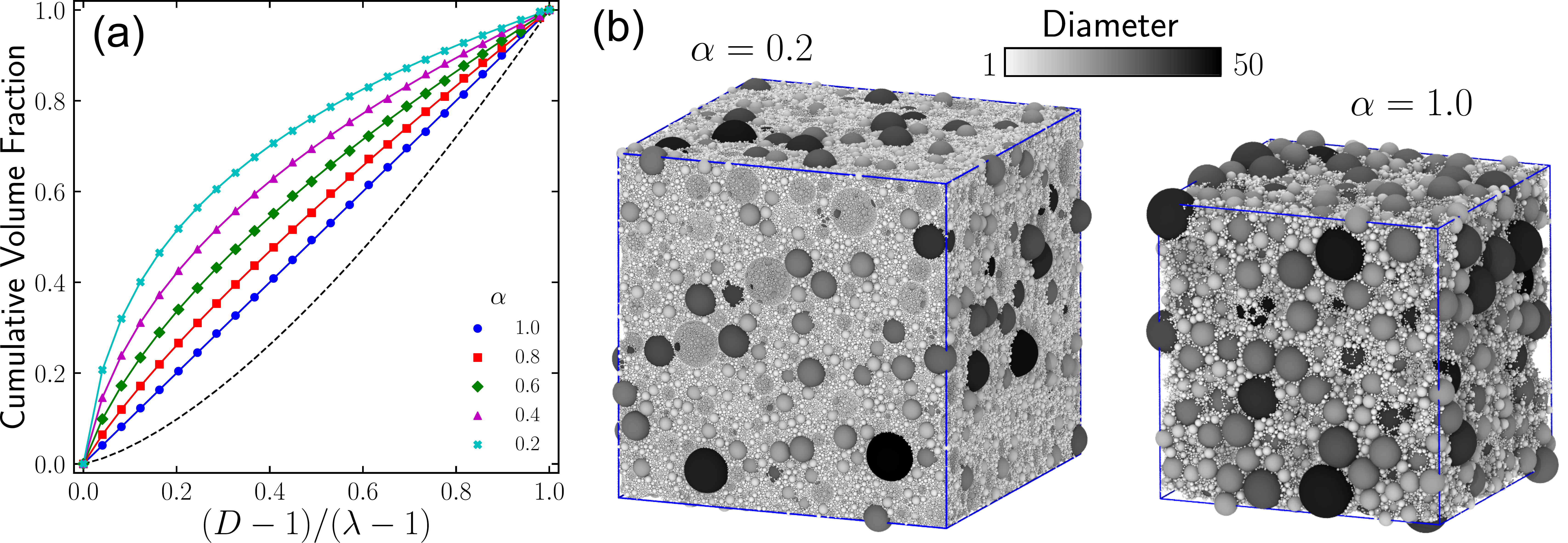}
\caption{(a) Cumulative volume fractions for select values of $\alpha$ and $\lambda = 50$. Points are computed from the particle distributions used in the simulations and lines are the analytic curves obtained from Eqn.~\eqref{eqn:cvf}. The dashed black line corresponds to $\alpha = 1.5$ and is included as an example case for $\alpha > 1.0$, for which the CVF is concave up. (b) Example packings obtained at applied pressure $p_{\rm a} = 10^{-6}$, shaded by particle diameter for the indicated $\alpha$. Both simulation boxes are triclinic but have small tilt factors. The overall number of particles for $\alpha = 0.2$ is $\sim\!16\times$ larger than for $\alpha = 1.0$, and the packed volume fractions are close to $\phi = 0.82$ and $\phi = 0.80$ for $\alpha = 0.2$ and $\alpha = 1.0$, respectively.
}
\label{fig:fig1}
\end{figure*}

\subsection{Contact model}
\label{sec:cmodel}
Spherical particle-based 3D DEM packing simulations were conducted using the GRANULAR package in LAMMPS~\cite{plimpton1995,silbert2001,thompson2022}.
The scope of this study is limited to frictionless, purely repulsive normal contacts, where particles interact via a damped Hookean pair potential penalizing overlap.
The normal force $\mathbf{F}_{\rm n}$ between contacting particles $i$ and $j$ with diameters $D_i$ and $D_j$ and separation $\mathbf{r}_{ij} = \mathbf{r}_i - \mathbf{r}_j$ is
\begin{equation}
    \mathbf{F}_{\rm n} = k_{\rm n}\delta\hat{\mathbf{n}} - M_{\rm eff}\gamma_{\rm n}\mathbf{v}_{\rm n},
    \label{eq:force}
\end{equation}
where $k_{\rm n} = 1$ is the Hookean spring constant, $\delta = (D_i + D_j)/2 - |\mathbf{r}_{ij}|$ is the overlap, $M_{\rm eff} = M_iM_j/(M_i+M_j)$ in terms of the particle masses $M_i$ and $M_j$, and $\gamma_n = 0.5$ is a damping coefficient, reflecting particle inelasticity.
The unit vector connecting the particle centers is $\hat{\mathbf{n}} = \mathbf{r}_{ij}/|\mathbf{r}_{ij}|$, and $\mathbf{v}_{\rm n}$ is the relative velocity of the two particles projected along $\hat{\mathbf{n}}$.
Note that in principle the net normal force in Eqn.~\eqref{eq:force} can be attractive, i.e., if the damping component is greater than the Hookean component when particles are moving apart.
An extra switching function is employed to set the magnitude of $\mathbf{F}_{\rm n}$ to zero if this condition occurs during the simulation.
We do not expect this formulation to cause any significant changes for slow compression simulations, but it may be an important consideration for high-rate deformation simulations, for instance.

The material density of individual particles is $\rho = 1$, such that particle masses $M_i$ are proportional to the particle volumes $V_i$ and given by $M_i = \rho V_i = \rho\pi D_i^3/6$.
The unit of length is the smallest particle diameter $D_{\rm min} = 1$, and the unit of pressure is $k_n/D_{\rm min}$; all lengths and pressures are given in terms of these quantities.
The simulation timestep is $\Delta t = 0.02\tau$, where $\tau = \sqrt{M_{\rm min}/k_n}$ with $M_{\rm min} \equiv \rho \pi/6$.

\subsection{Particle size distributions}
\label{sec:distro}
Particle sizes are represented using diameters $D$ and are distributed according to power-law distributions such that the probability of finding a particle with diameter between $D$ and $D+dD$ is $P(D)dD \propto D^{-\beta}dD$, where $\beta$ is the power-law distribution exponent.
Particle sizes are limited to a range $1 \leq D \leq \lambda$, where the parameter $\lambda$ denotes the maximum size ratio of the distribution.
Each system is required to have at least 10 particles with diameters larger than $0.95\lambda$, meaning that the total number of particles in each system depends upon both $\lambda$ and $\beta$.
In the geophysical literature, distributions of particle sizes are often given in terms of their fractal dimensions $d_{\rm f}$, meaning that the number of particles $N_D$ larger than size $D$ satisfies $N_D\sim D^{-d_{\rm f}}$~\cite{turcotte1986}.
For power-law particle size distributions,
\begin{equation}
    N_D \propto \int_{D}^\infty D'^{-\beta}dD' \propto \frac{D^{1-\beta}}{1-\beta},
    \label{eq:dfractal}
\end{equation}
for $\beta > 1$, so that $d_{\rm f} = \beta - 1$.

A central quantity of interest is the cumulative volume fraction (CVF), which gives the fraction of particle volume (and mass, since $\rho$ is constant) contained in particles smaller than a given size.
The CVF is easily obtained for power-law distributions with $\beta < 4$ as
\begin{equation} 
    \text{CVF} = \frac{\int_1^D D'^{3-\beta}dD'}{\int_1^\lambda D'^{3-\beta}dD'} = \frac{D^{4-\beta}-1}{\lambda^{4-\beta}-1}.
    \label{eqn:cvf}
\end{equation}
The CVF exponent, $\alpha$, is defined using Eqn.~\eqref{eqn:cvf} as $\alpha \equiv 4-\beta$, and gives the scaling of the CVF in the limits $D^\alpha \gg 1$ and $\lambda^\alpha \gg 1$ as $\text{CVF}\sim (D/\lambda)^\alpha$.
Note that $\alpha$ plays a similar role as the grain-size distribution (GSD) exponent $\eta$ in Refs.~\cite{oquendo2020,oquendo2021,oquendo2022}, i.e., GSD~$\sim [(D-1)/(\lambda-1)]^\eta$, but cannot be compared directly (except in the specific case $\alpha = \eta = 1.0$, corresponding to $\beta = 3$) as the underlying particle size distributions in Refs.~\cite{oquendo2020,oquendo2021,oquendo2022} are not power laws characterized by a single exponent.

This work mainly considers exponents in the range $3.0 \leq \beta \leq 3.8$, or $0.2 \leq \alpha \leq 1.0$, emblematic of soil comminution~\cite{turcotte1986,filgueira2006,ben-nun2010,minh2013,debono2020}.
Several of these CVFs are shown in Fig.~\ref{fig:fig1}(a) plotted against the reduced particle diameter $(D-1)/(\lambda-1)$.
A simple interpretation of Fig.~\ref{fig:fig1}(a) is that more than half of the particle volume (mass) is contained in particles with diameters smaller than the arithmetic mean diameter $(\lambda+1)/2$, or reduced particle diameter $1/2$, for $\alpha < 1.0$---the concave down curves in Fig.~\ref{fig:fig1}(a)---and more than half of the particle volume is contained in particles larger than $(\lambda+1)/2$ for $\alpha > 1.0$.
As demonstrated in Sec.~\ref{sec:charpack}, the range of values of $\alpha$ shown in Fig.~\ref{fig:fig1}(a) brackets the densest obtainable packing for $\lambda \gg 1$.
As $\alpha\rightarrow 0$ ($\beta\rightarrow 4$), the preponderance of particles have diameters close to $D = 1$ and the total count of particles rises sharply.
For example, for $\lambda = 50$, the $\alpha = 1.0$ system has $233,653$ particles while for $\alpha = 0.2$ there are $3,739,236$ particles.
The corresponding packings obtained at low applied pressure (see Section~\ref{sec:protocol}) for these two systems are shown in Fig.~\ref{fig:fig1}(b), rendered in OVITO~\cite{ovito}.

\subsection{Efficient multi-neighboring scheme}
\label{sec:multi}
To identify potentially interacting atoms/particles in MD and DEM packages, the most computationally efficient basic algorithm builds a neighbor list with all pairs of nearby particles using a spatial grid with a length scale set by the largest interaction cutoff.
This becomes impractical as the size disparity ratio $\lambda$ increases, as the same bin size is used for all particle pairs.
An alternative approach was implemented into LAMMPS by~\citet{veld2008}, which uses the smallest cutoff to set the bin size and to adjust how many bins are searched based on particle types \cite{veld2008}.
In LAMMPS, particle types are a discrete categorization used to set interaction parameters such as cutoffs for MD or friction coefficients for DEM.
While this method allows simulations to reach larger $\lambda$, it also becomes exceedingly expensive as $\lambda$ increases beyond $\sim\!10$.
To overcome this limitation, an improved algorithm was recently proposed by~\citet{ogarko2012} and initially modified for LAMMPS by~\citet{stratford2018} and~\citet{shire2021}.
This approach further tailored the neighbor list construction based on a particle's type to ensure that the computational cost of building a neighbor list does not grow faster with $\lambda$ than the force calculation. 

For this work, the implementation by~\citet{stratford2018} was expanded upon by fully integrating it with the LAMMPS codebase and releasing it in the public LAMMPS distribution \footnote{These capabilities are described in the documentation found at https://docs.lammps.org/neighbor.html. An example input script in.powerlaw is included with the LAMMPS distribution in the examples/multi subdirectory. It is a demo of shearing a 2D packing of particles with power-law-distributed sizes using neighbor list options similar to those leveraged in this work.}.
The method is generalized to support DEM by removing the use of particle types, since these are typically intended to describe material properties and not necessarily particle sizes.
Neighbor list construction can be tuned by pre-defining a set of diameter intervals irrespective of particle types, streamlining optimization of simulations.
The crux of the technique is that each particle searches for neighbors with diameters that fall in its own diameter interval, and in larger diameter intervals.
This approach takes advantage of the inherent asymmetry in the computational effort required to generate lists of neighbors centering on small particles as opposed to using large particles as the point of reference.
Previous work demonstrated that this method can be used to model jamming of frictionless and frictional bidisperse packings up to $\lambda = 40$ \cite{srivastava2021}.
Here, this methodology is applied to study packings of frictionless particles with a power-law distribution with $\lambda$ as large as 100, although larger $\lambda$ are feasible.
See the Appendix for arguments regarding the computational complexity of the algorithm and benchmark results.

\subsection{Constant-pressure packing protocol}

Packings are created via a constant-pressure protocol using the NPH ensemble implemented in LAMMPS~\cite{santos2020,srivastava2021}.
The symmetric applied pressure tensor, $\mathbf{P}_{\rm a}$, has the form $P_{\rm a,xx} = P_{\rm a,yy} = P_{\rm a,zz} = p_{\rm a}$ and all off-diagonal components are zero.
Here, $p_a$ is set to $10^{-6}$ to work in the limit of small particle overlaps; for context, in systems of monodisperse particles the typical fractional overlap is $\delta/D \sim p_a$ (in units of $D/k_n$).
The simulation box is fully periodic and initially cubic.
Under the constraint of no overlaps, i.e., there are no inter-particle forces at time $t = 0$, particles are randomly placed throughout the simulation box at low volume density.
The overall particle volume fraction, is defined as $\phi \equiv \left(\sum_{i} V_i\right)/V$, where $V$ is the instantaneous simulation box volume and the sum runs over all the particles in the system.

\label{sec:protocol}
\begin{figure}
\centering
\includegraphics[width=0.47\textwidth]{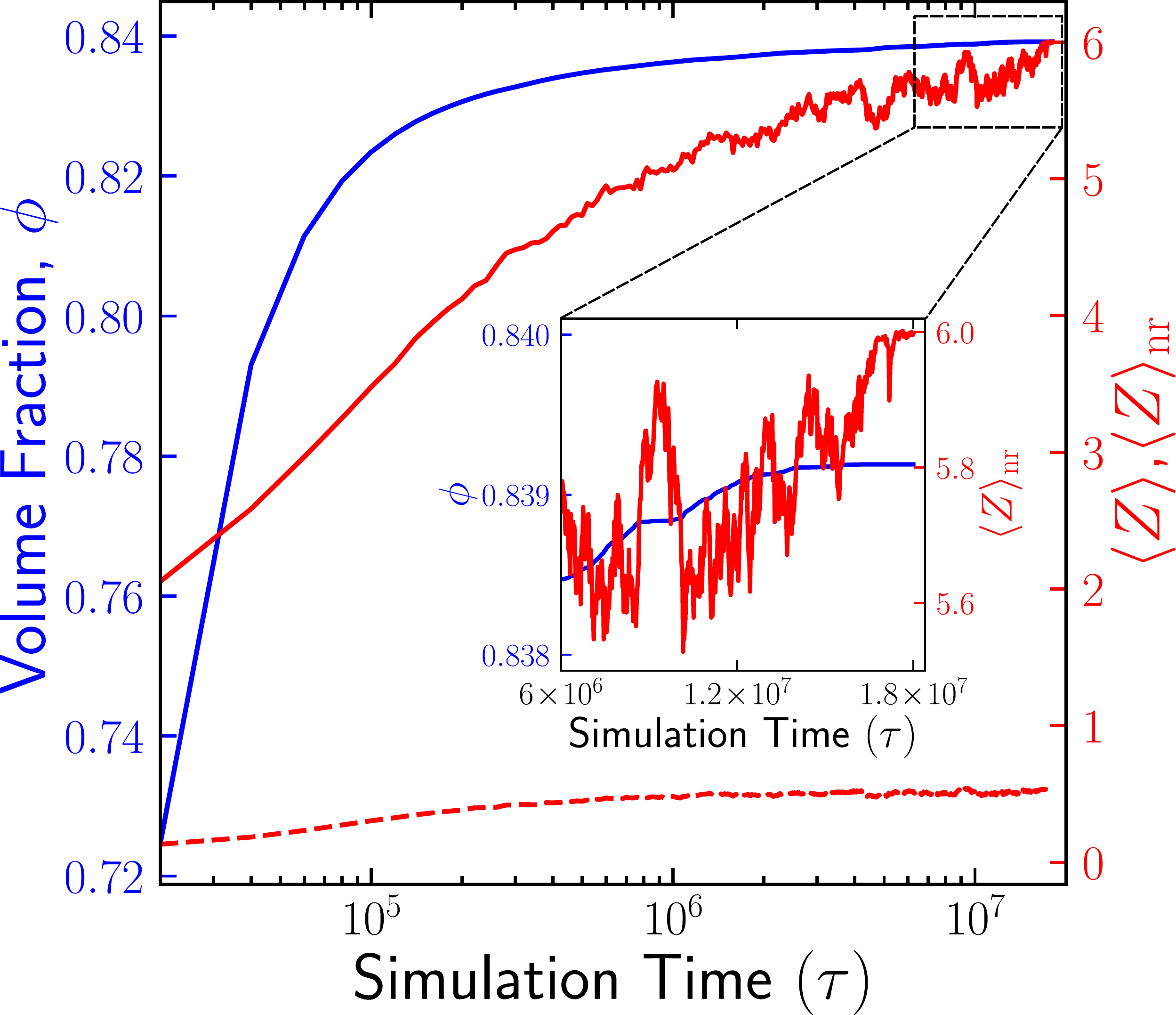}
\caption{Particle volume fraction $\phi$, overall mean coordination $\langle Z\rangle $ (dashed), and non-rattler mean coordination $\langle Z \rangle_{\rm nr}$ (solid), plotted against simulation time for $\lambda = 50$ and $\alpha = 0.7$. The inset magnifies the end of the simulation, at which time $\phi$ and $\langle Z \rangle_{\rm nr}$ cease evolving.}
\label{fig:fig2}
\end{figure}

During the simulation, the applied pressure compresses the simulation box and forces particles into contact; see Fig.~\ref{fig:fig2} for an illustration of the typical variation of $\phi$ with simulation time.
At the end of the simulation, the internal pressure tensor balances the applied pressure, giving $\mathbf{P}_{\rm int} = \mathbf{P}_{\rm a}$ within numerical tolerance.
While the simulation box is triclinic, the box tilt factors are typically small compared to the characteristic box side length.
The rate of compression is slow enough that the simulation box volume monotonically decreases until the system jams, and particle overlaps are much smaller than the particle diameters.

As the number of particles and the distribution of particle volumes vary substantially, we do not identify the final jammed state by using a fixed kinetic energy cutoff.
Rather, several criteria are used to determine when to stop jamming simulations.
In addition to the numerical equivalence of the final internal and applied pressure tensors, other quantities are also considered, including the evolution of $\phi$ and the mean number of contacts per particle, $\langle Z\rangle$, where $\langle\cdot\rangle$ refers to the average over all particles.
After jamming is achieved, $\phi$ and $\langle Z\rangle$ do not evolve in time and the total kinetic energy is small: the average kinetic energy per-particle is of order $10^{-13}$ or less.
For several of the largest systems---$\lambda = 50$ systems with $\beta \gtrsim 3.6$ ($\alpha \lesssim 0.4$)---a simulation time cutoff of at least $1.2 \times 10^7\tau$ and up to $\sim 4 \times 10^7\tau$ is employed out of computational necessity to stop simulations.
Most quantities extracted from the simulations, such as $\phi$ and $\langle Z\rangle$, evolve slowly if at all after such long run times (see Fig.~\ref{fig:fig2}).

In frictionless, monodisperse systems under vanishingly small pressure, the isostatic number of contacts per particle is $Z_{\rm iso} = 6$.
The number of excess contacts per particle, $\Delta Z = Z - Z_{\rm iso}$, grows systematically with pressure as $\sqrt{p_{\rm a}}$~\cite{ohern2003,liu2010,santos2020}.
For highly disperse packings, a large proportion of particles are rattlers, i.e, those particles participating in too few contact pairs to be mechanically stable.
Such particles substantially dilute the calculation of $\langle Z \rangle$, but a separate, more informative value, $\langle Z \rangle_{\rm nr}$, can be obtained by excluding rattler particles from the calculation.
The difference between the two measures is evident by comparing the two red curves in Fig.~\ref{fig:fig2}.
This issue will be discussed in greater detail in Sec.~\ref{sec:results}.

As a separate test of mechanical stability, we conducted several additional simulations to verify that the packings with the most extreme fractions of rattler particles  are still stable after removing rattlers (see Sec.~\ref{sec:charpack}). For example, for $\alpha = 1.0$ approximately $83\%$ and $97\%$ of all particles are rattlers for $\lambda = 20$ and $50$, respectively.
For several high-$\alpha$ values, the simulations were restarted after removing rattlers from the packed configurations and checked for re-convergence of the macroscopic quantities, including $\phi$ and $\langle Z\rangle_{\rm nr}$, under the same stress state.

A final useful metric, the Cundall parameter $C$, quantifies the typical unbalanced net per-particle force, per contact, in the system~\cite{oquendo2020}:
\begin{equation}
    C = \frac{\sum_p|\mathbf{F}_{\rm p}|}{\sum_c|\mathbf{F}_{\rm c}|},
\end{equation}
where the numerator sums over the magnitude of the net per-particle force and the denominator sums over the magnitude of each contact force.
Our results showed that a value of $C \lesssim 10^{-6}$ was indicative of a mechanically-stable system.

\section{Results}
\label{sec:results}
\subsection{Characterization of packings}
\label{sec:charpack}
\begin{figure}
\centering
\includegraphics[width=0.45\textwidth]{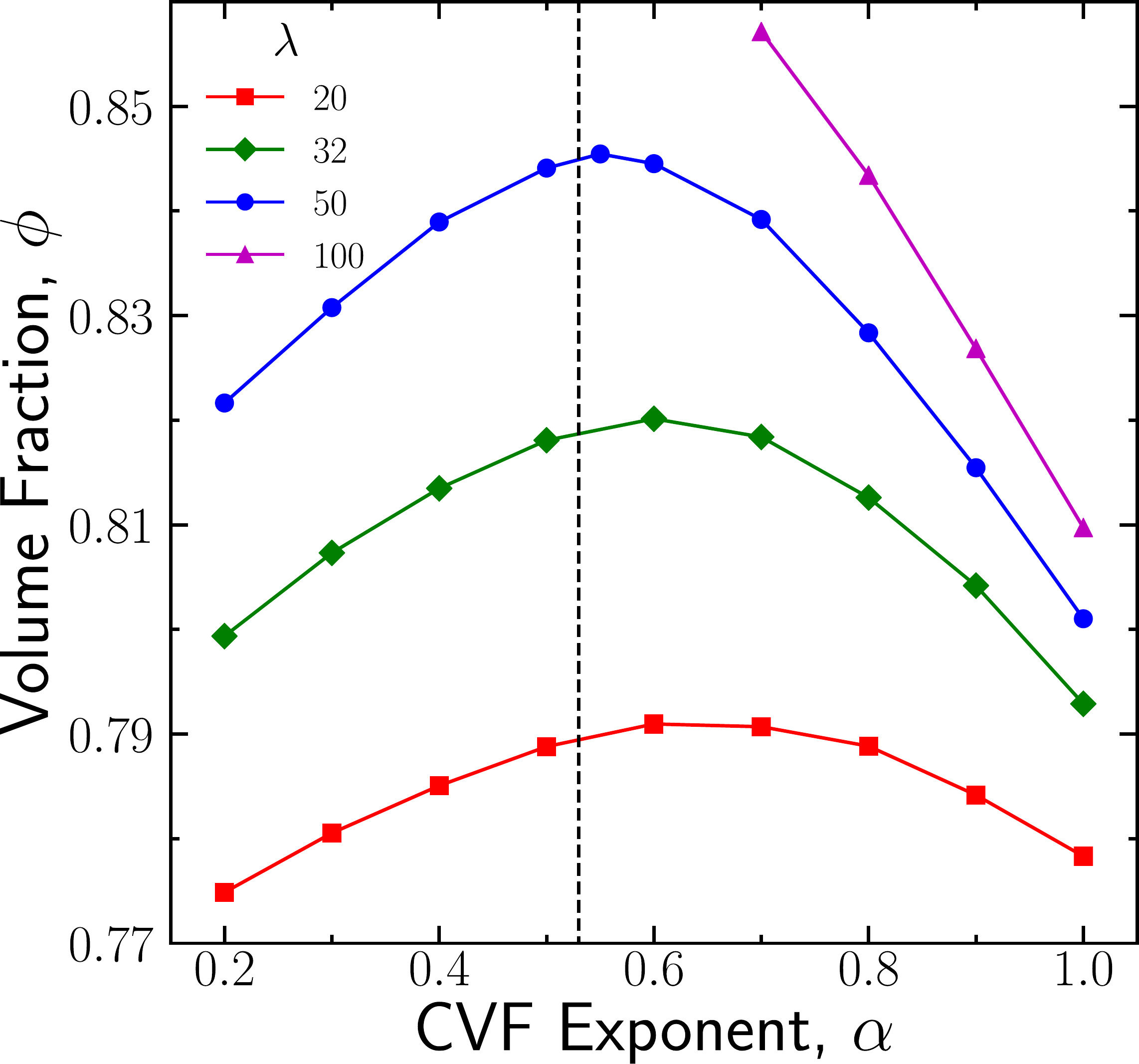}
\caption{Variation in the particle volume fraction $\phi$ of jammed packings plotted against $\alpha$ for the indicated values of the maximum particle size $\lambda$. The vertical dashed line corresponds to the CVF exponent $\alpha_{\rm A}\approx 0.53$ corresponding to the random Apollonian packing.
}
\label{fig:fig3}
\end{figure}
%
\begin{figure*}
\centering
\includegraphics[width=1\textwidth]{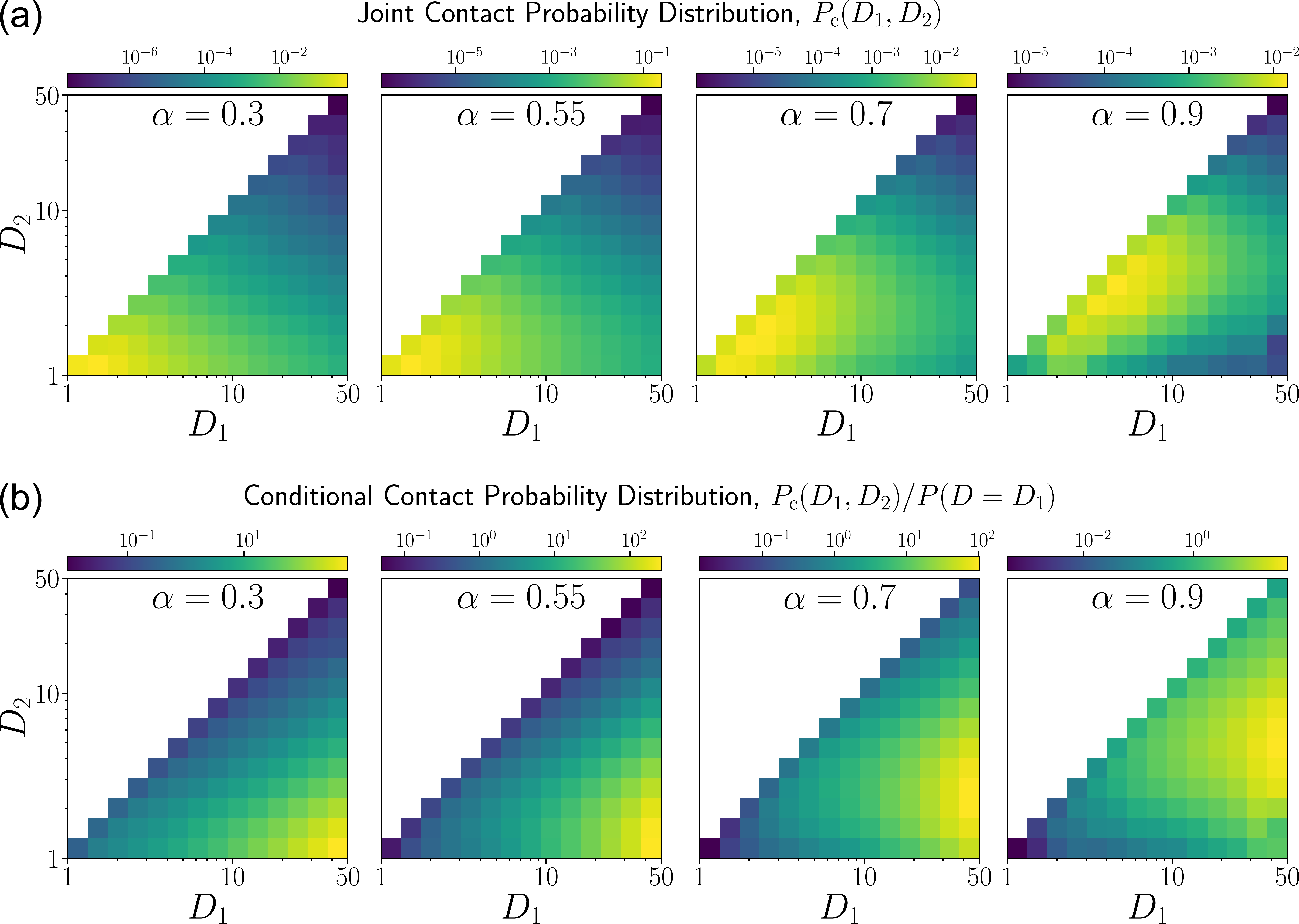}
\caption{(a) Joint contact probability distributions for particle diameter pairs with $D_1 \geq D_2$ for the indicated $\alpha$ and $\lambda = 50$. (b) Conditional contact probability distribution for the same set of systems.
}
\label{fig:fig4}
\end{figure*}

Using the constant-pressure packing protocol, packing volume fractions $\phi$ were obtained for different power-law particle size distributions.
Results comparing the variation of $\phi$ with $\alpha$ and $\lambda$ are shown in Fig.~\ref{fig:fig3}.
For each $\lambda$, $\phi$ is lowest at the endpoints of the range of $\alpha$ considered and peaks near the center of the range.
The peak shifts slightly to smaller $\alpha$ and becomes sharper with increasing $\lambda$.
The sharpening trend with increasing $\lambda$ is similar to behavior observed in bidisperse packings, for which increasing the particle size ratio changes $\phi$ at the Furnas peak from smoothly non-monotonic to cusped~\cite{prasad2017,srivastava2021}.
System size and packing equilibration constraints prevent us from exploring the entire range of $\alpha$ for $\lambda = 100$ to see if the sharpening trend persists in the power-law-distributed case.

The results in Fig.~\ref{fig:fig3} are consistent with the packing densities obtained by~\citet{oquendo2020,,oquendo2022}, who employed the Hertz contact model and volume-controlled isotropic compression.
Several other differences between our simulations and those described in Refs.~\cite{oquendo2020,oquendo2021,oquendo2022} warrant mentioning.
Based on the maximum fractional overlaps quoted in Ref.~\cite{oquendo2020}, the peak pressure reached in the simulations conducted in that work is estimated to be of order $p_{\rm a} \sim 10^{-4}$, or two orders of magnitude larger than in our simulations.
Test simulations we conducted for $\lambda = 20$ comparing the Hooke and Hertz contact models at identical $p_{\rm a} = 10^{-4}$ using the compression protocol described in Section~\ref{sec:protocol} showed that $\phi$ is $\sim\!1\%$ larger in Hertzian systems with all other variables kept constant, perhaps because the Hertz model does not penalize incipient particle overlap (the contact stiffness is zero at first contact).
Most importantly, as noted earlier, the GSD characterized by the exponent $\eta$ in Ref.~\cite{oquendo2020} does not correspond to an underlying single-exponent power-law size distribution, except in the case $\alpha = \eta = 1.0$.

Considering individual values of $\lambda$, it is possible to compare our CVFs with the GSDs in Ref.~\cite{oquendo2020} to estimate a value of $\alpha$ that compares most favorably with $\eta$.
For $\lambda = 32$---the largest $\lambda$ used in Ref.~\cite{oquendo2020}---a GSD with $\eta = 0.8$ approximates a CVF with $\alpha = 0.75$, $\eta = 0.6$ is similar to $\alpha = 0.5$, and $\eta = 0.4$ bears some resemblance to $\alpha = 0.2$, etc.
For $\eta\lesssim 0.3$, the approximated CVFs are logarithmic or have $\alpha < 0$.
In general, $\alpha$ values are smaller than the corresponding values of $\eta$.
Moreover, the GSD tends to exhibit a higher particle volume fraction contained by small particles, while the CVF and GSD exhibit similar scaling behaviors as $D\rightarrow\lambda$.
Despite this, the peak $\phi$ value obtained in this work is within $\sim\!5\%$ of that in Ref.~\cite{oquendo2020} and occurs at comparable $\alpha$ and $\eta$ values, for $\lambda = 32$.
Given the role of $\lambda$, the discrepancy between these results should reduce as the maximum particle size ratio increases.

Figure~\ref{fig:fig3} also indicates the CVF exponent of the power-law particle size distribution corresponding to the random Apollonian packing, $\alpha_{\rm A}\approx 0.53$~\cite{borkovec1994,anishchik1995,herrmann2003,varrato2011,oquendo2022}.
Interestingly, $\alpha_{\rm A}$ is quite close to the corresponding $\alpha$ used to obtain the peak $\phi$ for $\lambda = 50$, $\alpha = 0.55$, and as noted above the peak trends towards smaller $\alpha$ with increasing $\lambda$ (see also discussion in Ref.~\cite{oquendo2022}).
The power-law exponent of the Apollonian packing is conjectured to be the lower bound of exponents that result in full coverage obtained via geometric packing protocols~\cite{aste1996}.
The corresponding upper bound in 3D is $\beta = 4$ ($\alpha = 0$) ~\cite{aste1996}.
Our $\lambda = 50$ data can be compared with an extrapolation of data from Ref.~\cite{varrato2011}, which explored the physical fractal behavior of random Apollonian packings.
The densest packing we obtained gave $\phi\approx 0.85$, a packing significantly less dense than the corresponding Apollonian packing $\phi_{\rm A} \approx 0.93$.
Apollonian packings are created using particle insertion methods, which circumvent physical constraints on particle motion and have not been tested for mechanical stability, while in DEM simulations particles cannot move freely through constrictions/pores smaller than their diameters.
Thus, in jammed configurations, DEM-generated microstructures are expected to contain larger pores, resulting in overall looser packings than traditional Apollonian states.

The presence of the peak at intermediate $\alpha$ in Fig.~\ref{fig:fig3} suggests that obtaining the optimal packing density depends on balancing the abundance of the large-on-large particle contact pairs comprising the majority of the force-bearing backbone with the amount of small particles filling in the gaps between.
This supposition was suggested by~\citet{furnas1931} and is qualitatively supported by the snapshots shown in Fig.~\ref{fig:fig1}(b) for $\alpha = 1.0$ and $\alpha = 0.2$.
For the former, it is apparent that large particles regularly contact others of comparable size, but the relative scarcity of small particles available to populate the gaps results in a somewhat porous microstructure.
For the latter, contacts between large particles are rare because they are embedded in a sea of small particles.
Qualitatively similar behavior is observed in bidisperse packings, with high/low-$\alpha$ power-law disperse packings corresponding to low/high fractions of small particles in the bidisperse case~\cite{srivastava2021}.

To quantify this observation, the joint probability distributions for contacting pairs of particles are computed as determined by the particle diameters, $P_{\rm c}(D_1,D_2)$, as are the associated conditional contact probability distributions, $P_{\rm c}(D_1,D_2)/P(D = D_1)$, for $\lambda = 50$.
The results, ordered such that $D_1 \geq D_2$, are shown in Fig.~\ref{fig:fig4} for several $\alpha$.
From Fig.~\ref{fig:fig4}(a), the most probable pair of sizes for contact for $\alpha = 0.9$ occurs when both particles have intermediate diameters $\sim\!5-7$, close to the geometric mean of the maximum size ratio, $\sqrt{\lambda}$.
Conversely, for $\alpha = 0.3$ most contacts exist between pairs of particles with diameters near the smallest value.
Of course, since large particles are less abundant by construction, contacts between them make up a negligible fraction of the full set of contacts.
Figure~\ref{fig:fig4}(b) shows the conditional contact probability distributions, which better account for large particle scarcity.
Panel (b) shows that contacts including a particle with $D_1\sim \lambda = 50$ are more common than might be otherwise expected (because these particles have the greatest surface area), and further, that the diameter of its contact pair partner falls from $D_2\sim \sqrt{\lambda}$ for high $\alpha$ to $D_2 \sim 1$ for low $\alpha$.
This result underscores the increasing importance of small particles in stabilizing the packing as $\alpha$ decreases.
~\citet{minh2013} pointed out that changes in large particle connectivity from being large-particle-dominated to being small-particle-dominated may reduce the propensity of large particles to fracture.
If this hypothesis is correct, then our results indicate that low-$\alpha$ packings should be less susceptible to inter-particle fracture than high-$\alpha$ packings.

Figure~\ref{fig:fig4} does not show a strong signature distinguishing results for the densest packing obtained for $\lambda = 50$ ($\alpha  = 0.55$) from results for other $\alpha$.
From Fig.~\ref{fig:fig3}, the densest packing is not obtained for contact probability distributions that are large-large/intermediate pair dominated (e.g., $\alpha = 0.9$ in Fig.~\ref{fig:fig4}) or large-small pair dominated (e.g., $\alpha = 0.3$ in Fig.~\ref{fig:fig4}), but rather for $\alpha = 0.55$, which from Fig.~\ref{fig:fig4} lies somewhere in between these two extremes.
This observation aligns with the behavior of bidisperse packings, which reach optimal density through a saturation of large-large and small-small contact pairs~\cite{furnas1931,srivastava2021}.
%

\begin{figure*}
\centering
\includegraphics[width=\textwidth]{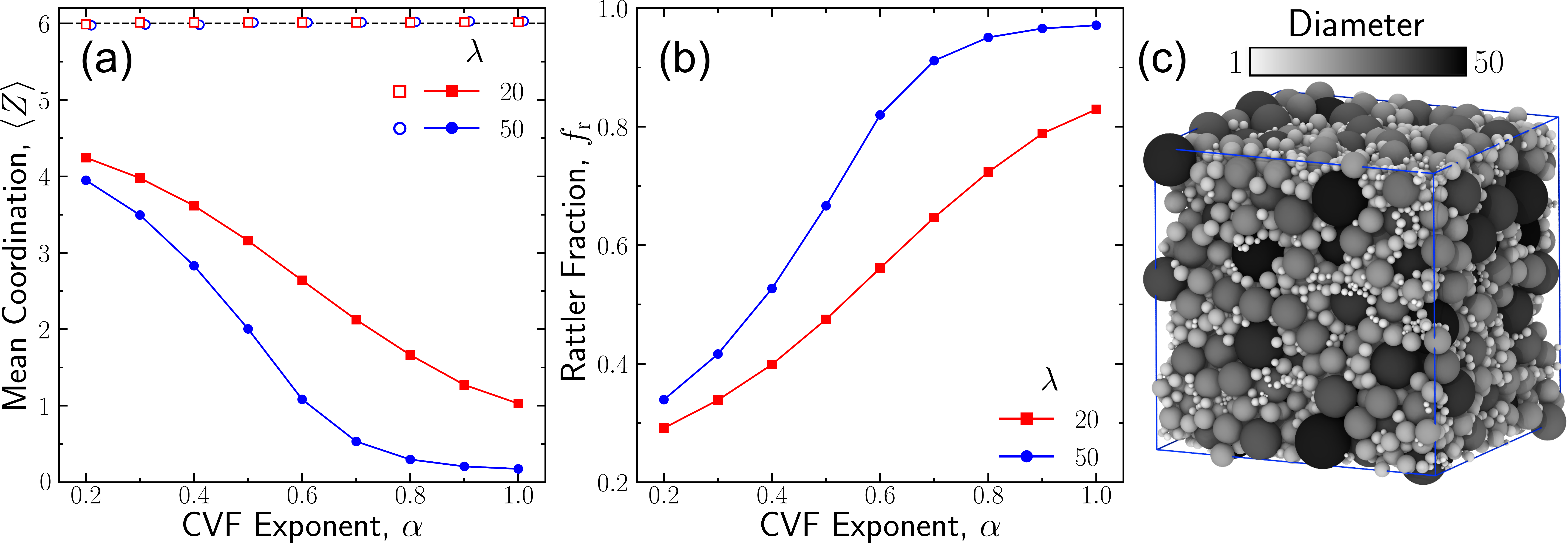}
\caption{(a) Mean coordination per particle including rattlers $\langle Z\rangle$ (filled symbols) and without rattlers $\langle Z\rangle_{\rm nr}$ (open symbols) for the indicated values of the maximum particle size, $\lambda$. (b) Fraction of rattler particles for the same systems. (c) Reproduced $\alpha = 1.0$ packing from Fig.~\ref{fig:fig1}(b) after removing rattlers.
}
\label{fig:fig5}
\end{figure*}

In power-law disperse systems, the number of contacts for each particle depends upon its size, but the behavior of the mean number of contacts or coordination $\langle Z \rangle$ is less clear.
As noted in Sec.~\ref{sec:protocol}, rattler particles strongly influence the calculation of $\langle Z\rangle$.
However, the overall value of $\langle Z\rangle$ and a rattlers-excluded value $\langle Z\rangle_{\rm nr}$ can be computed separately~\cite{roux2000}.
For the latter, rattler particles are identified by determining particles with fewer than $3$ contacts and removed from the list of contact pairs.
Note that each rattler removed decrements the total number of contacts for its contact pair partners, so the removal process is done iteratively~\cite{donev2004}.

After removing rattlers, the mean coordination is recomputed for the reduced contact list and the smaller set of non-rattler particles.
The results of these analyses for all particles and only non-rattler particles are shown in Fig.~\ref{fig:fig5}(a) as a function of $\alpha$ and for two separate values of $\lambda$.
Considering first the overall value of $\langle Z \rangle$, shown using filled symbols, the results are smaller than $Z_{\rm iso}$ for all $\alpha$ and both $\lambda$ values.
The largest $\langle Z \rangle$ is found for $\alpha = 0.2$, while $\langle Z \rangle$ is close to zero for $\alpha = 1.0$ for $\lambda = 50$.
After removing rattlers, Fig.~\ref{fig:fig5}(a) shows that $\langle Z \rangle_{\rm nr}$ is approximately equal to $Z_{\rm iso}$ for all $\alpha$ and $\lambda$ (shown as open symbols).
Since the value of $p_{\rm a}$ used in our simulations is small, the corresponding non-rattler value of $\Delta Z_{\rm nr}\equiv \langle Z\rangle_{\rm nr}-Z_{\rm iso}$ is likewise small but non-zero: typical values found in this work are $\Delta Z_{\rm nr} \approx 0.01-0.03$.

The results in Fig.~\ref{fig:fig5}(a) imply that the effect of rattler particles is significant for every system examined.
An oppositely related quantity to $\langle Z \rangle$ is the rattler fraction, $f_{\rm r}$, the fraction of all particles that are rattlers.
Figure~\ref{fig:fig5}(b) quantifies how $f_{\rm r}$ increases with $\alpha$.
In particular, for $\alpha = 1.0$ and $\lambda = 50$ only a few percent of particles are non-rattlers{; this system, with rattlers removed, is reproduced in Fig.~\ref{fig:fig5}(c)}.
Rattler fractions of comparable magnitude were also observed in Ref.~\cite{oquendo2020} for $\eta = \alpha = 1.0$.
Indeed, the trends shown in Fig.~\ref{fig:fig5} closely mirror those of Fig. 5 in Ref.~\cite{oquendo2020} and seem to be fairly universal.
Furthermore, the rattler fraction variation with $\alpha$ shown in Fig.~\ref{fig:fig5}(b) is reminiscent of the discontinuous jump in small particle rattler behavior observed at the Furnas peak in frictionless, bidisperse packings~\cite{furnas1931,srivastava2021}.
In Section~\ref{sec:norattlers}, the implications of these results on the distribution of particles that participate in the mechanical stability of the packing will be examined.

%
\begin{figure}
\centering
\includegraphics[width=0.45\textwidth]{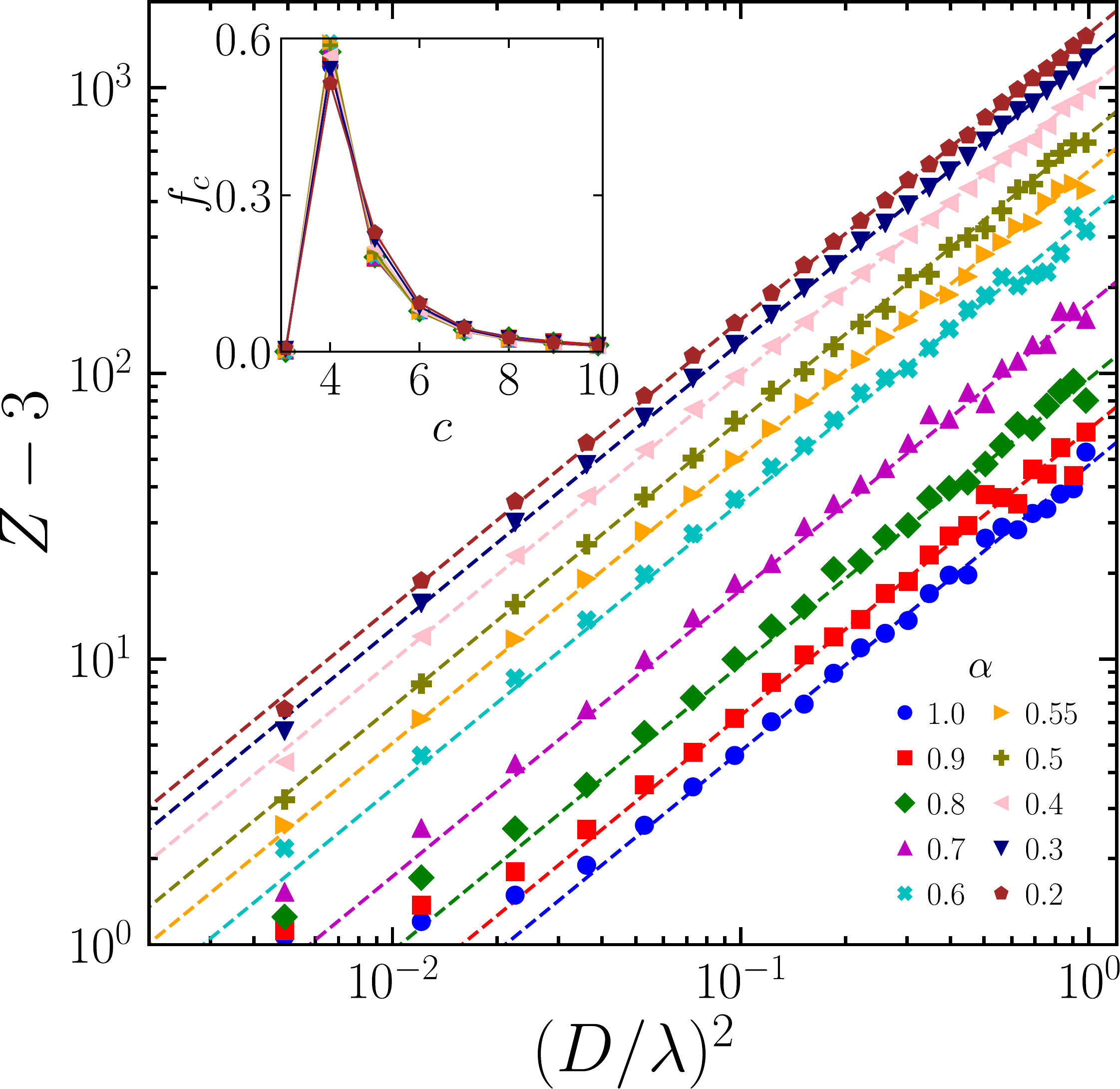}
\caption{Mean excess number of contacts per particle than are required for mechanical stability ($Z = 3$), as a function of $(D/\lambda)^2$ for the indicated $\alpha$ and $\lambda = 50$. Rattler particles are omitted from these results. Dashed lines are linear fits to the symbols of matching color. Inset: the fraction of non-rattler particles $f_c$ participating in exactly $c$ contacts.
}
\label{fig:fig6}
\end{figure}

While these results show that the mean non-rattler coordination is a constant, the exemplar packings depicted in Fig.~\ref{fig:fig1}(b) suggest that the number of contacts per particle, $Z(D)$, depends upon both particle diameter $D$ and the underlying distribution of particle sizes (see Ref.~\cite{minh2013} for a similar calculation for frictional particles).
The quantity $Z(D)$ is calculated by binning particles by size and computing the mean number of contacts per particle in each bin.
Rattler particles are excluded from this analysis to mitigate transient effects resulting from their short-lived participation in contact pairs.
Results for $\lambda = 50$ are shown in Fig.~\ref{fig:fig6}.
The figure demonstrates the scaling of $Z-3$, i.e., the mean excess number of contacts over the three contacts required for mechanical stability, plotted against $(D/\lambda)^2$; plotted in this way, the relationship is linear and corresponding linear fits to data for large $(D/\lambda)^2$ were computed (dashed lines).
Note that $(D/\lambda)^2 = 0.01$ corresponds to $D = 5$.
The relationship $Z-3\sim D^2$ exhibits the same $D^2$ scaling with diameter as the particle surface area, and represents a slightly faster scaling than was observed in Ref.~\cite{minh2013} for smaller, frictional particles.
It is clear from the linear fits in Fig.~\ref{fig:fig6} that the prefactor decreases with increasing $\alpha$, an intuitive result given that fewer intermediate and large particles than small particles can be placed in the available solid angle of any central particle~\cite{corwin2010,danisch2010}.
From Fig.~\ref{fig:fig4}(b), this exclusion of solid angle inherent to large-large contact pairs has strongest significance for high $\alpha$, resulting in the lowest overall maximum per-particle contact count.
Linear fits also worked for both  $\lambda = 20$ and $\lambda = 100$ (not shown), though the prefactors generally depended upon $\lambda$ for $\alpha \lesssim 0.8$.

The results in Fig.~\ref{fig:fig6} are striking given that $\langle Z\rangle_{\rm nr}$ is approximately six, while $Z(D)$ for the largest particles is at least an order of magnitude larger.
The inset of Fig.~\ref{fig:fig6} shows the fraction of non-rattler particles $f_c$ participating in exactly $c$ contacts, which is peaked at $c = 4$ and essentially independent of $\alpha$.
This low coordination value is responsible for the deviations away from linear scaling for small $(D/\lambda)^2$.
Note that this analysis distinguishes between the discrete contact count $c$ and the bin-averaged quantity $Z(D)$.
Similar results for $f_c$ were reported in Ref.~\cite{mutabaruka2019} for smaller size dispersity and different underlying particle size distributions. 
The range $3\leq c\leq 10$ encompasses between 90--95\% of all non-rattler particles for each $\alpha$ but only accounts for roughly 80\% of $\langle Z \rangle_{\rm nr} = \sum_c cf_c$.
The remaining contributions to $\langle Z \rangle_{\rm nr}$ come from the high, but rare, contact participation counts of large particles.

\subsection{Non-rattler particle distributions}
\label{sec:norattlers}
%
\begin{figure}
\centering
\includegraphics[width=0.45\textwidth]{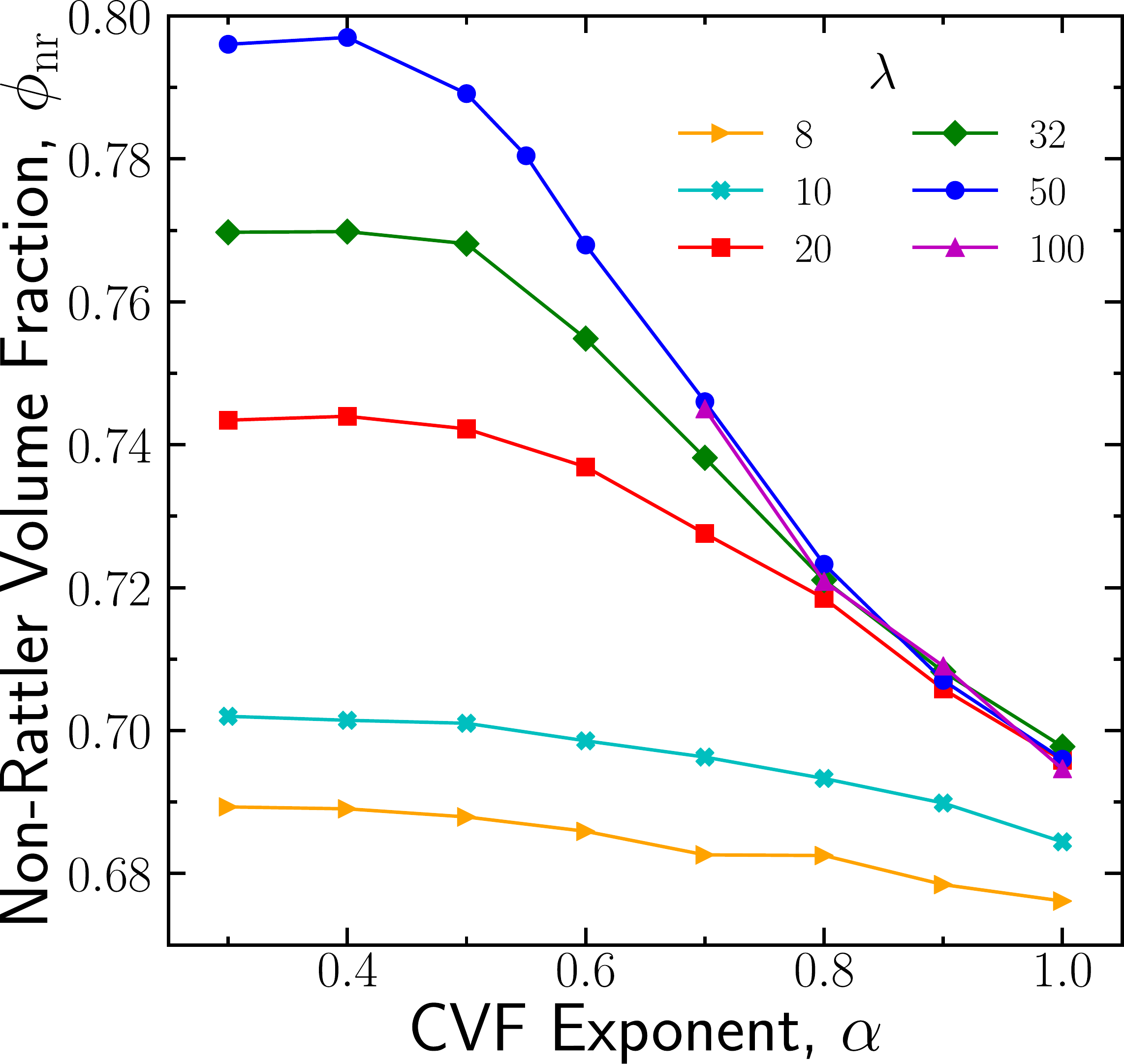}
\caption{Volume fraction $\phi_{\rm nr}$ contributed exclusively by non-rattler particles for the indicated $\lambda$.
}
\label{fig:fig7}
\end{figure}

Rattler particles contribute to the overall volume fraction and density of the jammed packing, but have no bearing on its mechanical stability.
Large particles are crucial to the force-bearing network, while sizable fractions of the small particles are rattlers.
This suggests that the input particle size distribution differs from the distribution of particles in the force-bearing network.
This section examines how the shape of the input particle size distribution $P(D)$ dictates the resultant distribution of non-rattler particles, $P_{\rm nr}(D)$, and is motivated by considering the volume fraction contributed solely by non-rattler particles $\phi_{\rm nr}$, shown in Fig.~\ref{fig:fig7}.
This measure is akin to the mechanical void ratio in the geophysical literature~\cite{otsubo2016,liu2021}.
In contrast to the clear dependence of the overall particle volume fraction $\phi$ on $\lambda$ shown earlier in Fig.~\ref{fig:fig3}, Fig.~\ref{fig:fig7} shows that $\phi_{\rm nr}$ is \textit{independent} of $\lambda$ (for $\lambda \gtrsim 20$) at high $\alpha$.
While not shown here, results for $\alpha = 1.5$ and $\alpha = 2.0$ also collapsed for $\lambda \geq 8$.
In this shallow power-law limit ($\beta\rightarrow0$, $\alpha\rightarrow 4.0$), $\lambda$ must become irrelevant.
As $\alpha$ grows, $\phi_{\rm nr}$ gradually approaches the equivalent monodisperse packing value with rattlers removed,  $\phi_{\rm nr}^{\rm mono}$; for the constant-pressure protocol and averaged over five realizations each with $10^5$ monodisperse particles, we obtained $\phi_{\rm nr}^{\rm mono} \approx 0.629$, compared to the overall packing density $\phi^{\rm mono}\approx 0.639$.

Several interesting trends are apparent for $\alpha\rightarrow 0.3$ in Fig.~\ref{fig:fig7}.
First, $\phi_{\rm nr}$ levels off at low $\alpha$, and second, the plateau values steadily increase with $\lambda$.
The $\lambda \geq 20$ data in Fig.~\ref{fig:fig7} suggest that increasing $\lambda$ may shift the collapse of $\phi_{\rm nr}$ to progressively smaller $\alpha$.
Since reducing $\alpha$ corresponds to increasing the relative abundance of small particles compared to large particles, a plateau in $\phi_{\rm nr}$ implies that there may be diminishing returns to adding \textit{more} small particles as most become rattlers.
However, adding smaller and smaller particles, i.e., increasing $\lambda$, does lead to denser force-bearing networks in the plateau regime.
It is interesting that while $\alpha\sim 0.55-0.6$ lead to the densest overall packings, the force-bearing components of such packings are less dense than those for smaller $\alpha$.
In most cases, the volume fraction lost when rattlers are removed, $\phi - \phi_{\rm nr}$, is smaller than 0.1, with the largest shifts occurring for $\alpha = 1.0$.
This is an indication that while rattlers may constitute a large fraction of the total number of particles, they typically only account for a small fraction of the total particle volume.
%
\begin{figure}
\centering
\includegraphics[width=0.45\textwidth]{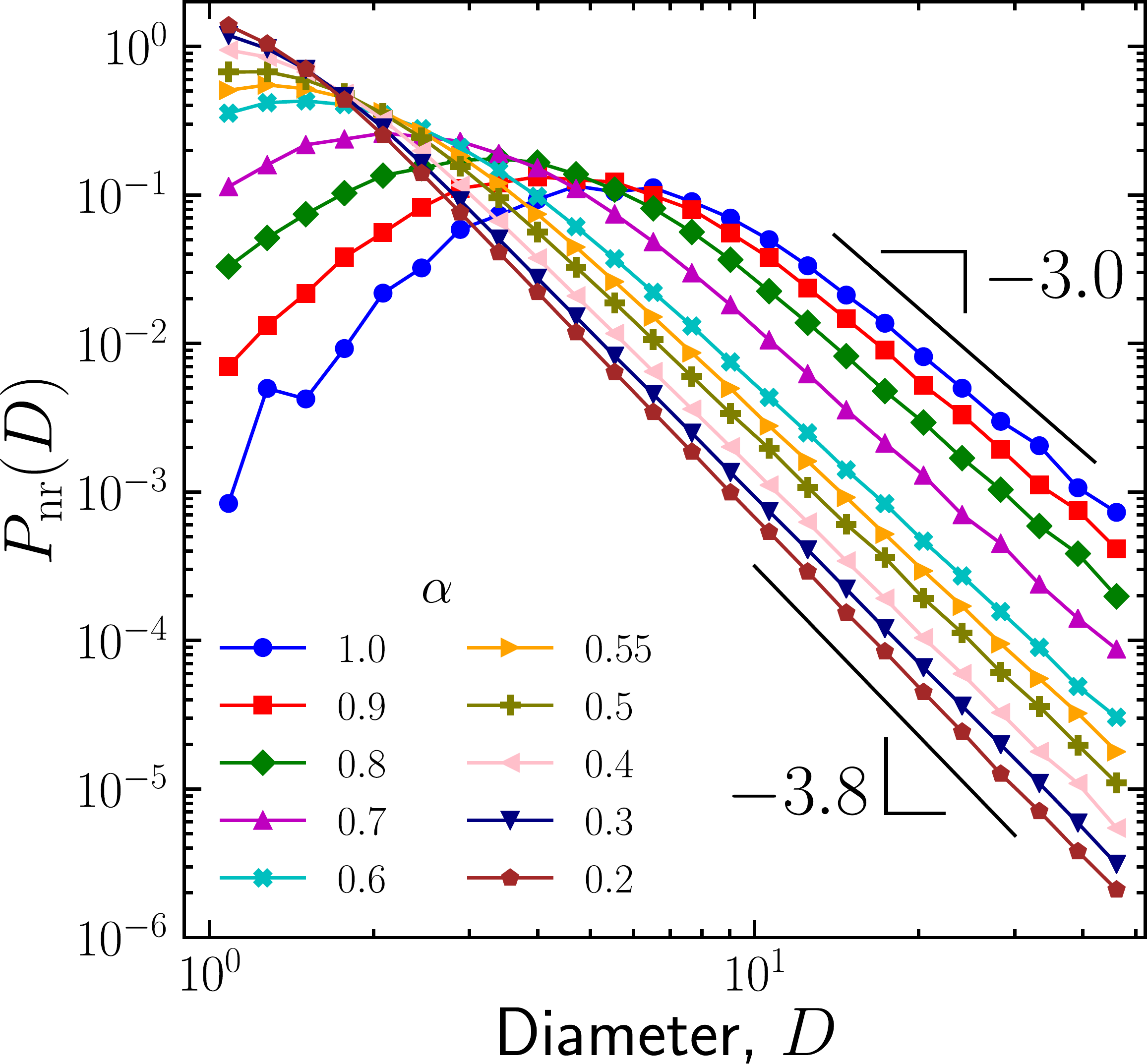}
\caption{Distributions of non-rattler particles for the indicated $\alpha$ and $\lambda = 50$. For each curve, the power-law behavior at large $D$ is unchanged from the initial particle size distribution, with exponent $\beta = \alpha-4$. The limiting cases of such power-law scalings are indicated.
}
\label{fig:fig8}
\end{figure}
%
\begin{figure*}
\centering
\includegraphics[width=1\textwidth]{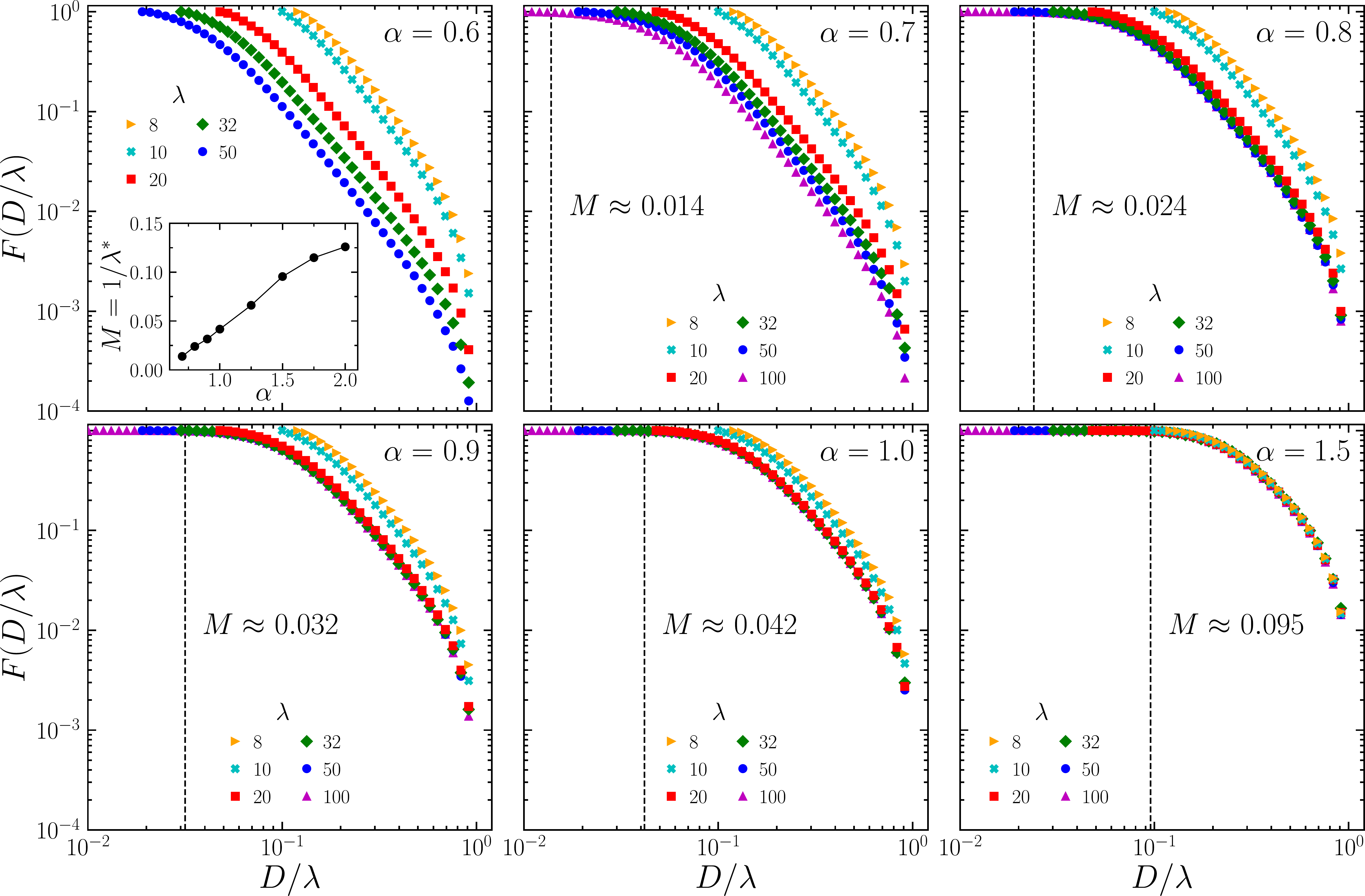}
\caption{Fraction of non-rattler particles with diameters larger than $D/\lambda$. The dashed black lines and printed labels $M$ mark the value of $D/\lambda$ for which $F(D/\lambda) = 0.99$ as estimated from the $\lambda = 100$ data. The inset to the upper left panel plots computed values of $M$ for $0.7\leq \alpha\leq 2.0$ that were obtained for $\lambda = 100$. Note that $M$ was computed for several additional values of $\alpha$ for which we omit plots of $F(D/\lambda)$ data.
}
\label{fig:fig9}
\end{figure*}

The removal of rattler particles permits the identification of the non-rattler particle size distributions $P_{\rm nr}$, the distribution governing the force-bearing backbone.
Results for $P_{\rm nr}$ obtained using $\lambda = 50$ are shown in Fig.~\ref{fig:fig8}.
For each $\alpha$, the tail of the distribution maintains its original power-law character, while the probability of retaining small particles is reduced, with substantial dependence on $\alpha$.
As $\alpha\rightarrow 1.0$, the most probable remaining particle diameters are close to $\sqrt{\lambda}$ (e.g., refer to the $\alpha = 1.0$ packing image in Fig.~\ref{fig:fig5}(c)), while for $\alpha\rightarrow 0.2$ it is clear that the smallest non-rattler particles remain the most probable.
Indeed, aside from the exponent of the power-law tail, only a small amplitude change for $D\sim 1$ differentiates $\alpha = 0.2$ from $\alpha = 0.55$, the densest overall packing.
The results shown in Fig.~\ref{fig:fig8}, taken together with the context given by Fig.~\ref{fig:fig5}(b) and Fig.~\ref{fig:fig7}, imply that high-$\alpha$ packings do not derive mechanical stability from small particles, while particles of all sizes are necessary to stabilize low-$\alpha$ packings.

The collapse of $\phi_{\rm nr}$ with increasing $\lambda$ for high $\alpha$ indicates that in such cases the underlying distributions of non-rattler particles should have similarities.
Rather than considering the non-rattler particle size distributions themselves, this analysis focuses on the fraction of non-rattler particles $F(x)$ that are larger than $x$.
This quantity has the advantages that it varies monotonically from 1 to 0, and Eq.~\eqref{eq:dfractal} dictates how it should scale with $D$ away from the endpoints.
To compare data for different $\lambda$ on an equal basis, the particle diameters are normalized by $\lambda$ such that the scaled diameters fall in the domain $1/\lambda\leq x = D/\lambda \leq 1$.
Results for $F(D/\lambda)$ are shown in Fig.~\ref{fig:fig9} for several $\alpha$ and a wide range of $\lambda$.
For each data set, $F(D/\lambda)$ is unity until the smallest non-rattler particle is encountered, beyond which $F(D/\lambda)$ drops to 0 in a manner that exhibits the expected power-law behavior over narrow ranges of $D/\lambda$ and accelerates as $D/\lambda\rightarrow 1$.
The power-law regime broadens as $\alpha$ increases, consistent with the results depicted in Fig.~\ref{fig:fig8}.

The most striking result shown in Fig.~\ref{fig:fig9} is that $F(D/\lambda)$ is identical for all $\lambda$ for $\alpha = 1.5$, signifying that the shape of the non-rattler particle size distribution is constant with respect to increases in $\lambda$ beyond $\lambda = 8$.
Similar results were obtained for larger $\alpha$ (not shown).
However, as $\alpha$ falls to 1.0 and lower, data for the smallest $\lambda$ increasingly deviate from the other curves until all data sets are clearly distinct for $\alpha \lesssim 0.7$.
In cases where $F(D/\lambda)$ collapses, the constant value of $D/\lambda$ determining the onset of $F<1$ means that there is a reduction of the effective width of the non-rattler particle size distribution, i.e., $\lambda \rightarrow \lambda^*$, with $\lambda^* \leq \lambda$.
This results from the removal of rattler particles with diameters smaller than a threshold $D^*_{\rm min} > D_{\rm min} = 1$.
In what follows, the onset value is referred to as the magnification, $M\equiv 1/\lambda^* = D^*_{\rm min}/\lambda$, which can be estimated for different $\alpha$.
Specifically, using the $\lambda = 100$ data shown in Fig.~\ref{fig:fig9}, $M$ is determined by extracting the onset value of $D/\lambda$ for which $F<0.99$.
Note that there is ambiguity in the precise determination of $M$ based on the threshold $F$ value---for example, for $\alpha = 1.5$, both the $\lambda = 8$ and $\lambda = 10$ data nearly collapse on top of the larger $\lambda$ data despite being smaller than the nominal $\lambda^*\approx 10.5$.
However, our tests showed that changes in the estimated magnifications are sharper for higher $F$ thresholds, and the estimates of $M$ we obtained for our threshold choice are sufficient for the discussion here.

The estimated values of $M$ are given in the corresponding panels of Fig.~\ref{fig:fig9} for $\alpha \geq 0.7$ and indicated with dashed black lines.
In addition, an inset plotting $M(\alpha)$ estimated from the $\lambda = 100$ data is shown in the $\alpha = 0.6$ panel, and includes data for several $\alpha$ that are not shown in Fig.~\ref{fig:fig9}.
In essence, the definition of $M$ permits us to determine a criterion, given by $1/\lambda \lesssim M$, for which the non-rattler particle size distribution is independent of $\lambda$.
For systems satisfying this criterion, particles with $D < D^*_{\rm min}$ are almost always rattlers.
Moreover, $D^*_{\rm min}$ replaces $D_{\rm min} = 1$ as the unit of length of the system; as an intensive quantity, $\phi_{\rm nr}$ is also independent of $\lambda$ when $M$ is constant.
Note that since the tail of the non-rattler particle size distribution is unchanged from the original power law, it is still possible to collapse $F$ for each $\alpha$, provided that a $\lambda$-dependent rescaling factor is used.
In such cases, however, $\phi_{\rm nr}$ does not collapse.

To help contextualize these results, the two separate limits of $M = 1$ and $M = 1/\lambda$ can be defined.
The former is relevant for high-$\alpha$ systems that approach the limit of monodisperse systems, which is defined by a singular length scale set by the particle diameter.
The latter occurs in cases where the smallest particles are necessary for ensuring mechanical stability of the packing.
From the trend depicted in the inset of Fig.~\ref{fig:fig9}, the $M = 1$ limit is likely slowly approached for $\alpha > 2$.

The results in Fig.~\ref{fig:fig9} show that the force-bearing component of the packing is invariant with respect to changes in $\lambda$ provided that the $\alpha$-dependent scale $\lambda^*$ is exceeded.
From this standpoint, no additional benefit is gained by adding successively smaller particles once $\lambda \geq \lambda^*$.
However, the overall properties and structure of the packing, including the packing density, still depend on the full particle size distribution.
Indeed, the relatively low density of the force-bearing component suggests that the rattler particles play a significant role in determining the final configuration, perhaps by restricting the intermediate configurations that the force-bearing component can adopt.
Intriguing avenues for possible future study include successive, repeated jamming and removal of rattler particles to isolate the limiting particle size distribution, and in designing particle size distributions for use in constructing the densest possible packings that can be obtained via compaction protocols.
In addition, DEM simulations of frictional and/or cohesive particles with large size dispersity have not yet been systematically performed, but are crucial to connecting simulation results with real-world applications.

\section{Conclusion}
We performed large-scale 3D DEM simulations to study the packing properties of power-law disperse spherical particles.
This work considered a wide range of power-law particle size distributions, varying the range of particle sizes and the exponents characterizing the power laws.
To our knowledge, we have simulated and studied particle size ratios larger than any other 3D DEM study to date.

At fixed particle size ratio, the results showed that the densest overall packings were obtained for power-law particle size distributions that achieved mechanical stability while balancing contacts between pairs of large-large and large-intermediate particles with pairs of small-small and large-small particles.
Distributions containing too many large particles do not need small particles for mechanical stability, and so most small particles were rattlers.
Conversely, distributions with too many small particles produce packings that are dominated by large-small contact pairs, and so do not generate contacts between pairs of particles with diameters in the intermediate size classes.
Further, despite the strong dependence of the mean coordination and rattler fraction on the CVF exponent, the mean coordination of non-rattler particles was close to the isostatic value, while the mean number of contacts per non-rattler particle scaled quadratically with particle diameter.

Considering only non-rattler particles, volume fractions of non-rattler particles for input distributions with high $\alpha$ were independent of size dispersity for $\lambda$ larger than an $\alpha$-dependent cutoff value $\lambda^*$, while for low $\alpha$ the non-rattler volume fraction was insensitive to $\alpha$.
In the former case, the fraction of non-rattler particles with normalized diameters larger than $D/\lambda$ was independent of $\lambda$, provided that $\lambda \geq \lambda^*$.
This result signifies a separate effective length scale of the force-bearing network, as determined by $\alpha$.
For the latter case, the results indicated that increasing the \textit{proportion} of small particles has little effect on the force-bearing component of the packing, while adding \textit{smaller} particles tended to improve both the overall and non-rattler packing densities.

The results presented here provide insight into the internal microstructure of large size dispersity particle packings and broaden our understanding of the relationship between features of the overall packing and its force-bearing backbone.
In turn, the understanding gleaned from studying these systems may provide a pathway to optimizing the properties, mechanical and otherwise, of designed particle packings.

\section{Acknowledgements}
The authors thank Andrew P. Santos for helpful discussions and Kevin Stratford, Tom Shire, and Kevin Hanley for providing an initial implementation of their disperse neighboring technique in LAMMPS and Steve Plimpton and Axel Kholmeyer for their assistance in expanding this implementation and merging it into LAMMPS for public use.
I. S. acknowledges support from the U.S. Department of Energy (DOE), Office of Science, Office of Advanced Scientific Computing Research, Applied Mathematics Program under Contract No. DE-AC02- 05CH11231. 
This work was performed at the Center for Integrated Nanotechnologies, a U.S. DOE and Office of Basic Energy Sciences user facility.
Sandia National Laboratories is a multi-mission laboratory managed and operated by National Technology and Engineering Solutions of Sandia, LLC., a wholly owned subsidiary of Honeywell International, Inc., for the U.S. Department of Energy National Nuclear Security Administration under contract DE-NA0003525.
The views expressed in the article do not necessarily represent the views of the U.S. Department of Energy or the United States Government.

\appendix*
\section{Multi-neighboring details and performance}
\label{appendix}
To simulate systems with large disparity ratios $\lambda$ in particle sizes, we leveraged contact detection algorithms optimized for highly disperse systems. 
In this Appendix, we focus on dense, homogeneous systems and derive the leading order computational costs in $\lambda$ for three methods.
First, we describe a typical approach for monodisperse systems~\cite{plimpton1995,thompson2022}.
Next, we consider two additional methods designed for highly disperse systems: an older method by ~\citet{veld2008} and a more recent technique by~\citet{ogarko2012} and~\citet{krijgsman2014} that was extended and ported to LAMMPS for the first time by~\citet{stratford2018} and~\citet{shire2021}.
Our arguments will highlight a fundamental advantage of the newest method.
A brief description of each approach is included although further details can be found in their respective references.
Finally, we briefly describe changes in our implementation of the newest method in LAMMPS aimed at continuous particle size distributions and provide some benchmark data for the power-law-distributed systems studied in the main text.

\begin{figure}[t]
\centering
\includegraphics[width=0.45\textwidth]{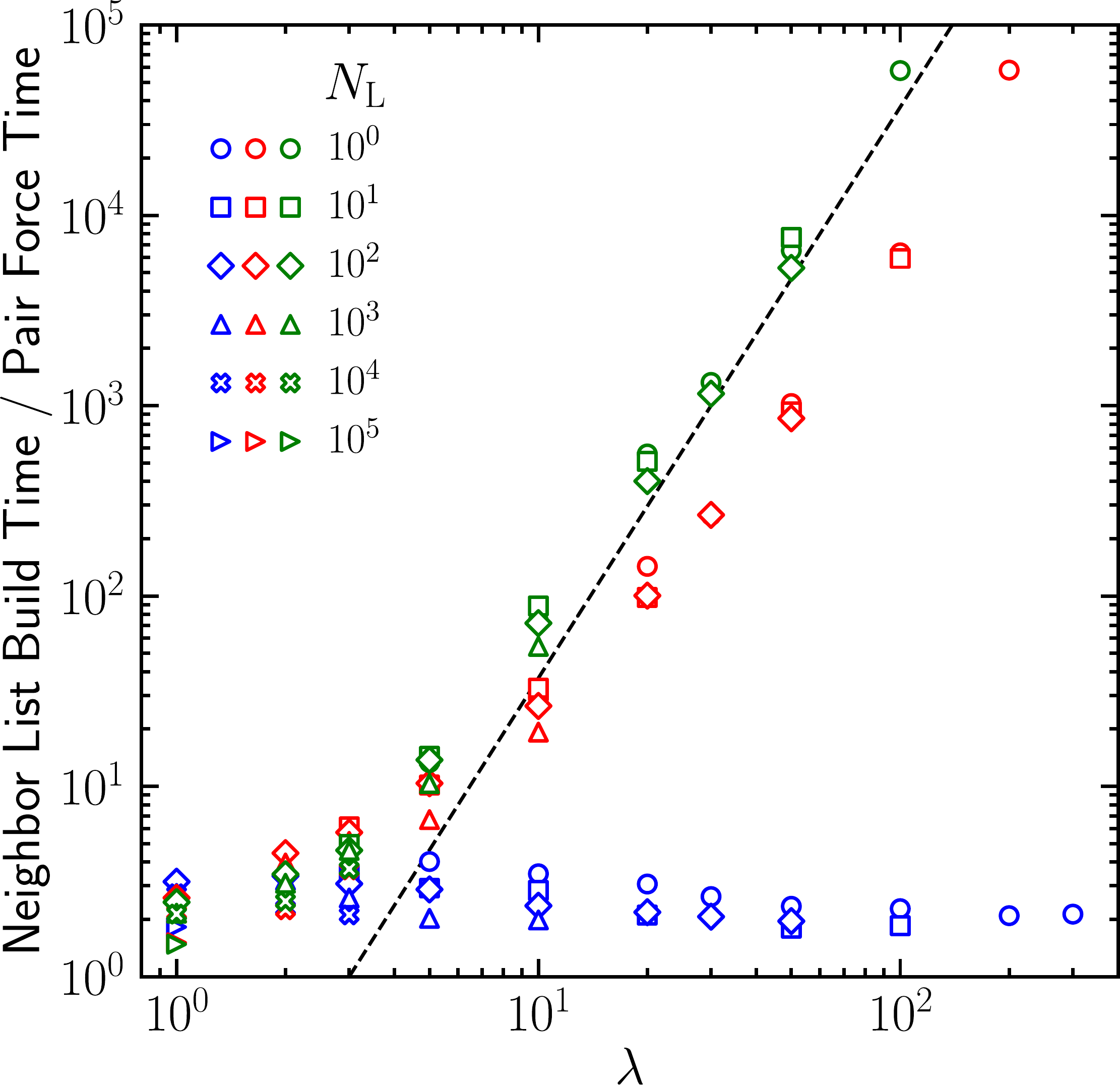}
	\caption{Relative cost of constructing a neighbor list compared to a single force evaluation for 3D binary packings near jamming with $f_{\rm S} = 0.5$ and the indicated values of $\lambda$ and $N_{\rm L}$ using the {\it default} (green), {\it multi/old} (red), and {\it multi} (blue) algorithms. Tests were run on a single processor. The dashed line represents $\lambda^3$ scaling. Note that typically there are many force evaluations between neighbor list builds in LAMMPS and this reported value does not represent a typical ratio over many timesteps. For perspective at $\lambda = 200$, building a single neighbor list took less than $9$ seconds using {\it multi} but over 72 hours using {\it multi/old}. Builds were impractically long to measure using the {\it default} method at $\lambda > 100$ or {\it multi/old} at $\lambda > 200$.
} 
\label{fig:neighbor_scaling}
\end{figure}

\begin{figure}[t]
\centering
\includegraphics[width=0.45\textwidth]{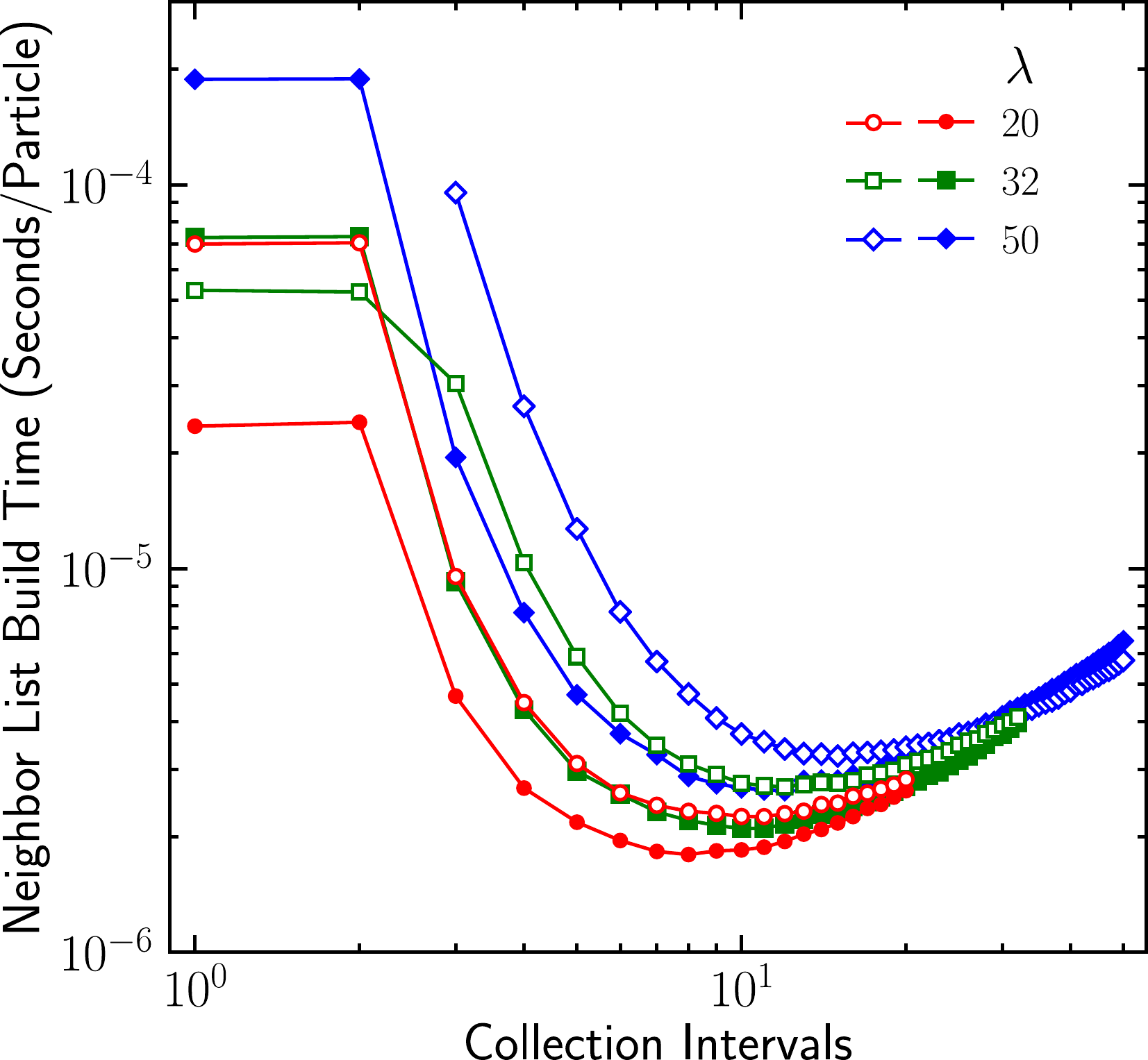}
	\caption{Time in seconds to build a single neighbor list per particle as a function of the number of collection intervals for jammed packings of power-law-distributed grains for the indicated $\lambda$. Calculations were performed on a single processor. Open symbols: $\beta = 3.0$ $(\alpha = 1.0)$; filled symbols: $\beta = 3.6$ $(\alpha = 0.4)$.} 
	\label{fig:benchmark_power}
\end{figure}
Before calculating contact forces, particle-based simulations often construct a Verlet neighbor list which contains all pairs of interacting particles~\cite{plimpton1995, thompson2022}.
In LAMMPS, a link-cell method is used where particles are spatially binned onto a grid with a bin size $\Delta$ in a process that takes $\mathcal{O}(N_\mathrm{T})$ time where $N_\mathrm{T}$ is the total number of particles.
Particles use this binning to efficiently generate a list of potential neighbors consisting of particles within their own bin and in other nearby bins that are within the interaction distance.
This set of bins that need to be searched is known as a stencil.
A distance is then only calculated between these candidate neighbors as opposed to all $\sim N_\mathrm{T}^2$ pairs of particles in the system.
In the default algorithm in LAMMPS, labeled {\it default}, $\Delta$ is approximately half of the maximum interaction distance such that the stencil only includes a small number of adjacent bins \footnote{Some aspects of neighbor list construction in LAMMPS, such as the addition of a skin to the interaction distance and the existence of full and half stencils, are not described in this Appendix as they do not affect the underlying scaling of costs with $\lambda$.}.
This method is very efficient for nearly monodisperse systems.

To illustrate how this algorithm fails at large $\lambda$, we consider a $d$-dimensional, bidisperse packing of $N_{\rm L}$ large particles and $N_{\rm S}$ small particles with diameters $D_{\rm L}$ and $D_{\rm S}$, respectively, and a volume fraction of small particles $f_{\rm S} \equiv N_{\rm S} D_{\rm S}^d/(N_{\rm S} D_{\rm S}^d + N_{\rm L} D_{\rm L}^d)$. 
As in the main text, one can treat $D_{\rm S} = D_{\rm min} = 1$ such that $D_{\rm L}/D_{\rm S} = \lambda$.
We mainly focus on a fixed volume fraction $0 < f_{\rm S} < 1$ such that $N_{\rm S} = N_{\rm L} \lambda^d f_{\rm S}/(1 - f_{\rm S})$ and $N_\mathrm{T} \approx N_{\rm S} \gg N_{\rm L}$ at large $\lambda$, but other cases are briefly discussed at the end of the Appendix.
In the {\it default} method, the bin size $\Delta$ is set by $D_{\rm L}$ such that each bin contains $N_\mathrm{particles/bin} \sim D_{\rm L}^d/D_{\rm S}^d = \lambda^d$ particles and the stencil only contains a finite number $N_\mathrm{bins/stencil}$ of nearby bins.
Therefore, there are $N_\mathrm{T} N_\mathrm{particles/bin} N_\mathrm{bins/stencil}$ candidate neighbors and calculating their distances is $\mathcal{O}(N_{\rm L} \lambda^{2d})$.
This cost is greater than binning particles and dominates neighbor list construction for large $\lambda$.

To put this scaling in context, one can compare it to the total number of contacts in the system or the computational cost of calculating forces.
In a jammed system, each small particle can only have a finite number of contacts independent of $\lambda$ while each large particle can have up to $\sim D_{\rm L}^{d-1}/D_{\rm S}^{d-1} = \lambda^{d-1}$ contacts.
Therefore, the total number of contacts in the system scales as $N_{\rm S} + N_{\rm L} \lambda^{d-1} \sim N_{\rm S} \sim N_{\rm L} \lambda^d$.
This implies that the cost to build the neighbor list using the {\it default} method dominates the total simulation time and simulations become prohibitively expensive with increasing $\lambda$.
This disparity is seen in Fig. \ref{fig:neighbor_scaling} for $d = 3$ and $f_{\rm S} = 0.5$  where the ratio of time to construct the neighbor list versus the time to calculate forces grows as $\lambda^3$.

To reduce costs, an alternate algorithm was implemented in LAMMPS by~\citet{veld2008}.
We refer to this algorithm by its current name in LAMMPS, {\it multi/old}.
The {\it multi/old} method adjusts spatial binning based on a particle's type, a categorization of particles that is used to set interactions parameters including the distance cutoff for non-DEM particles.
In this approach, the size of a bin is set by the smallest interaction length $D_{\rm S}$ such that $N_\mathrm{particles/bin}$ is constant and no longer grows with $\lambda$.
Therefore, a different stencil is needed for each combination of particle types.
These stencils extend out to the order of $D_{\rm L}/D_{\rm S} = \lambda$ bins for large-large pairs, $(D_{\rm L} + D_{\rm S})/D_{\rm S} \sim \lambda$ bins for large-small pairs, and a constant number of bins for small-small pairs.
While this method reportedly accelerates highly disperse simulations up to a factor of 100 for $\lambda = 20$~\cite{veld2008}, it does not address the fundamental scaling with $\lambda$ due to the search for large-small neighbors.
Each small particle searches $N_\mathrm{bins/stencil} \sim (D_{\rm L}/D_{\rm S})^d = \lambda^d$ bins within the large-small stencil to find potential large neighbors, where most bins will not contain a large particle, such that $\mathcal{O}(N_{\rm S} \lambda^d)$ or $\mathcal{O} (N_{\rm L} \lambda^{2d})$ operations are still performed.
This scaling, with a smaller prefactor than {\it default}, is seen in Fig. \ref{fig:neighbor_scaling}.
Practically, we find simulations become intractable around $\lambda$ of order 10.

To address this shortcoming, an additional twist described in Refs. \cite{ogarko2012, stratford2018} uses a hierarchy of binning grids, one for each particle type in the initial LAMMPS implementation by~\citet{stratford2018}.
This method is referred to as {\it multi}, reflecting its current name in LAMMPS, and includes a separate binning grid for each particle type with a bin size $\Delta$ set by the same-type interaction distance.
In a binary system, $\Delta$ is set by $D_\mathrm{S}$ for small particles and $D_\mathrm{L}$ for large particles such that $N_\mathrm{particles/bin}$ does not depend on $\lambda$, similar to {\it multi/old}.
The key difference is that each particle looks for same-type neighbors using its own set of bins while only small particles look for large neighbors using the large bins.
Large particles do not search for small neighbors.
Therefore, $N_\mathrm{bins/stencil}$ is also independent of $\lambda$ and construction costs are proportional to $N_\mathrm{T} N_\mathrm{particles/bin} N_\mathrm{bins/stencil} \sim N_\mathrm{T} \sim N_{\rm L} \lambda^d$, equivalent to the force calculation.
This scaling is demonstrated in Fig.~\ref{fig:neighbor_scaling} where the time to construct a neighbor list normalized by the time to calculate forces has no significant dependence on $\lambda$ up to $\lambda = 300$ for {\it multi}.
The only factor preventing simulations at larger $\lambda$ was the growing number of particles in the system, already reaching $N_{\rm T} = 27$ million at $\lambda = 300$. 

Here, we considered the case of fixed $f_{\rm S}$ although one could consider a value of $f_{\rm S}$ that grows or shrinks with $\lambda$.
If $f_{\rm S}$ grew with $\lambda$ approaching the limit of unity, then $N_\mathrm{T} \sim N_{\rm S}$ would grow faster than $\lambda^d$, e.g., as $\lambda^{d+\varepsilon}$. 
For {\it multi}, $N_\mathrm{particles/bin}$ and $N_\mathrm{bins/stencil}$ would still both be finite such that neighbor costs would still grow as $N_\mathrm{T} \sim \lambda^{d+\varepsilon}$.
This increase in cost would mirror a increase in cost to calculate forces simply due to having more particles.
In contrast for {\it default}, $N_\mathrm{particles/bin}$ would still scale as $\lambda^d$ such that neighbor list construction would be $\mathcal{O}(N_\mathrm{T} \lambda^d)$ or $\mathcal{O}(\lambda^{2d + \varepsilon})$ and still dominate simulation time.
In the opposite limit, one could consider a shrinking $f_{\rm S}$ exemplified by the extreme case of $N_{\rm S} = 1$ where large particles effectively only have a finite number of contacts with other large particles.
Again, $N_\mathrm{particles/bin}$ and $N_\mathrm{bins/stencil}$ would be finite in {\it multi} but now only large-large look ups would be relevant such that building a neighbor list would be $\mathcal{O}(N_{\rm L})$, equivalent to calculating forces.
For {\it default}, the one small particle would be irrelevant and building the neighbor list would resemble the $\mathcal{O}(N_{\rm L})$ process for a purely monodisperse system, identical to the scaling of {\it multi} although with reduced overhead.
Therefore, in all of these cases, the scaling of {\it multi} will always match that of a force evaluation and will either scale better than or equivalent to {\it default}, although prefactors depend on the specific system.

For continuous particle size distributions, increasing the number of collection intervals with increasing $\lambda$ generally improves performance as particles are binned using a bin size $\Delta$ closer to their actual diameter.
These savings grow until the overhead of having additional collection intervals exceeds the benefit.
This is seen in Fig.~\ref{fig:benchmark_power} where the cost of neighbor list construction is plotted as a function of the number of linearly-spaced intervals for a variety of jammed systems with different power-law size distributions.
The optimal number of bins and their spacing ultimately depends on the specific system.
It is worth noting that as the optimal number of collection intervals continues to grow with $\lambda$, the computational costs of building a neighbor list could begin growing faster than the force evaluations with $\lambda$ as more binning grids need to be created and searched.
Practically, we have not yet found this to be a limitation, particularly since the optimal number of bins only reaches $\sim 10$ for the systems considered here.
Similar studies on the optimization of power-law-distributed particle sizes were also performed in Ref.~\cite{krijgsman2014}.

For this work, we expanded the original {\it multi} implementation of~\citet{stratford2018} to fully integrate it with LAMMPS and added the method to the public distribution of LAMMPS.
As the particle type in LAMMPS is intended to represent material properties such as moduli or friction coefficients and not necessarily the size of a particle, we also generalized their implementation and provided the option for users to select an arbitrary set of particle size intervals or collections independent of particle types.
This approach more closely reflects the original discussion of the method by~\citet{ogarko2012} and helps streamline optimization of neighbor list construction for continuous particle size distributions.

\bibliography{bib}

\begin{thebibliography}{49}%
\makeatletter
\providecommand \@ifxundefined [1]{%
 \@ifx{#1\undefined}
}%
\providecommand \@ifnum [1]{%
 \ifnum #1\expandafter \@firstoftwo
 \else \expandafter \@secondoftwo
 \fi
}%
\providecommand \@ifx [1]{%
 \ifx #1\expandafter \@firstoftwo
 \else \expandafter \@secondoftwo
 \fi
}%
\providecommand \natexlab [1]{#1}%
\providecommand \enquote  [1]{``#1''}%
\providecommand \bibnamefont  [1]{#1}%
\providecommand \bibfnamefont [1]{#1}%
\providecommand \citenamefont [1]{#1}%
\providecommand \href@noop [0]{\@secondoftwo}%
\providecommand \href [0]{\begingroup \@sanitize@url \@href}%
\providecommand \@href[1]{\@@startlink{#1}\@@href}%
\providecommand \@@href[1]{\endgroup#1\@@endlink}%
\providecommand \@sanitize@url [0]{\catcode `\\12\catcode `\$12\catcode
  `\&12\catcode `\#12\catcode `\^12\catcode `\_12\catcode `\%12\relax}%
\providecommand \@@startlink[1]{}%
\providecommand \@@endlink[0]{}%
\providecommand \url  [0]{\begingroup\@sanitize@url \@url }%
\providecommand \@url [1]{\endgroup\@href {#1}{\urlprefix }}%
\providecommand \urlprefix  [0]{URL }%
\providecommand \Eprint [0]{\href }%
\providecommand \doibase [0]{https://doi.org/}%
\providecommand \selectlanguage [0]{\@gobble}%
\providecommand \bibinfo  [0]{\@secondoftwo}%
\providecommand \bibfield  [0]{\@secondoftwo}%
\providecommand \translation [1]{[#1]}%
\providecommand \BibitemOpen [0]{}%
\providecommand \bibitemStop [0]{}%
\providecommand \bibitemNoStop [0]{.\EOS\space}%
\providecommand \EOS [0]{\spacefactor3000\relax}%
\providecommand \BibitemShut  [1]{\csname bibitem#1\endcsname}%
\let\auto@bib@innerbib\@empty
\bibitem [{\citenamefont {Fuller}\ and\ \citenamefont
  {Thompson}(1907)}]{fuller1907}%
  \BibitemOpen
  \bibfield  {author} {\bibinfo {author} {\bibfnamefont {W.~B.}\ \bibnamefont
  {Fuller}}\ and\ \bibinfo {author} {\bibfnamefont {S.~E.}\ \bibnamefont
  {Thompson}},\ }\href@noop {} {\bibfield  {journal} {\bibinfo  {journal}
  {Trans. Am. Soc. Civ. Eng.}\ }\textbf {\bibinfo {volume} {59}},\ \bibinfo
  {pages} {67} (\bibinfo {year} {1907})}\BibitemShut {NoStop}%
\bibitem [{\citenamefont {Andreasen}\ and\ \citenamefont
  {Andersen}(1930)}]{andreasen1930}%
  \BibitemOpen
  \bibfield  {author} {\bibinfo {author} {\bibfnamefont {A.~H.~M.}\
  \bibnamefont {Andreasen}}\ and\ \bibinfo {author} {\bibfnamefont
  {J.}~\bibnamefont {Andersen}},\ }\href@noop {} {\bibfield  {journal}
  {\bibinfo  {journal} {Kolloid-Z.}\ }\textbf {\bibinfo {volume} {50}}
  (\bibinfo {year} {1930})}\BibitemShut {NoStop}%
\bibitem [{\citenamefont {Furnas}(1931)}]{furnas1931}%
  \BibitemOpen
  \bibfield  {author} {\bibinfo {author} {\bibfnamefont {C.~C.}\ \bibnamefont
  {Furnas}},\ }\href {https://doi.org/10.1021/ie50261a017} {\bibfield
  {journal} {\bibinfo  {journal} {Ind. Eng. Chem. Res.}\ }\textbf {\bibinfo
  {volume} {23}},\ \bibinfo {pages} {1052} (\bibinfo {year}
  {1931})}\BibitemShut {NoStop}%
\bibitem [{\citenamefont {Turcotte}(1986)}]{turcotte1986}%
  \BibitemOpen
  \bibfield  {author} {\bibinfo {author} {\bibfnamefont {D.~L.}\ \bibnamefont
  {Turcotte}},\ }\href {https://doi.org/10.1029/JB091iB02p01921} {\bibfield
  {journal} {\bibinfo  {journal} {J. Geophys. Res.}\ }\textbf {\bibinfo
  {volume} {91}},\ \bibinfo {pages} {1921} (\bibinfo {year}
  {1986})}\BibitemShut {NoStop}%
\bibitem [{\citenamefont {Langston}\ \emph {et~al.}(1997)\citenamefont
  {Langston}, \citenamefont {Nikitidis}, \citenamefont {T{\"{u}}z{\"{u}}n},
  \citenamefont {Heyes},\ and\ \citenamefont {Spyrou}}]{langston1997}%
  \BibitemOpen
  \bibfield  {author} {\bibinfo {author} {\bibfnamefont {P.~A.}\ \bibnamefont
  {Langston}}, \bibinfo {author} {\bibfnamefont {M.~S.}\ \bibnamefont
  {Nikitidis}}, \bibinfo {author} {\bibfnamefont {U.}~\bibnamefont
  {T{\"{u}}z{\"{u}}n}}, \bibinfo {author} {\bibfnamefont {D.~M.}\ \bibnamefont
  {Heyes}},\ and\ \bibinfo {author} {\bibfnamefont {N.~M.}\ \bibnamefont
  {Spyrou}},\ }\href {https://doi.org/10.1016/S0032-5910(97)03288-9} {\bibfield
   {journal} {\bibinfo  {journal} {Powder Technol.}\ }\textbf {\bibinfo
  {volume} {94}},\ \bibinfo {pages} {59} (\bibinfo {year} {1997})}\BibitemShut
  {NoStop}%
\bibitem [{\citenamefont {Liu}\ \emph {et~al.}(2019)\citenamefont {Liu},
  \citenamefont {Dong}, \citenamefont {Tang}, \citenamefont {Krishnan},
  \citenamefont {Sant},\ and\ \citenamefont {Bauchy}}]{liu2019}%
  \BibitemOpen
  \bibfield  {author} {\bibinfo {author} {\bibfnamefont {H.}~\bibnamefont
  {Liu}}, \bibinfo {author} {\bibfnamefont {S.}~\bibnamefont {Dong}}, \bibinfo
  {author} {\bibfnamefont {L.}~\bibnamefont {Tang}}, \bibinfo {author}
  {\bibfnamefont {N.~M.}\ \bibnamefont {Krishnan}}, \bibinfo {author}
  {\bibfnamefont {G.}~\bibnamefont {Sant}},\ and\ \bibinfo {author}
  {\bibfnamefont {M.}~\bibnamefont {Bauchy}},\ }\href
  {https://doi.org/10.1016/j.jmps.2018.10.003} {\bibfield  {journal} {\bibinfo
  {journal} {J. Mech. Phys. Solids}\ }\textbf {\bibinfo {volume} {122}},\
  \bibinfo {pages} {555} (\bibinfo {year} {2019})}\BibitemShut {NoStop}%
\bibitem [{\citenamefont {O'Hern}\ \emph {et~al.}(2003)\citenamefont {O'Hern},
  \citenamefont {Silbert}, \citenamefont {Liu},\ and\ \citenamefont
  {Nagel}}]{ohern2003}%
  \BibitemOpen
  \bibfield  {author} {\bibinfo {author} {\bibfnamefont {C.~S.}\ \bibnamefont
  {O'Hern}}, \bibinfo {author} {\bibfnamefont {L.~E.}\ \bibnamefont {Silbert}},
  \bibinfo {author} {\bibfnamefont {A.~J.}\ \bibnamefont {Liu}},\ and\ \bibinfo
  {author} {\bibfnamefont {S.~R.}\ \bibnamefont {Nagel}},\ }\href
  {https://doi.org/10.1103/PhysRevE.68.011306} {\bibfield  {journal} {\bibinfo
  {journal} {Phys. Rev. E}\ }\textbf {\bibinfo {volume} {68}},\ \bibinfo
  {pages} {011306} (\bibinfo {year} {2003})}\BibitemShut {NoStop}%
\bibitem [{\citenamefont {Farr}\ and\ \citenamefont {Groot}(2009)}]{farr2009}%
  \BibitemOpen
  \bibfield  {author} {\bibinfo {author} {\bibfnamefont {R.~S.}\ \bibnamefont
  {Farr}}\ and\ \bibinfo {author} {\bibfnamefont {R.~D.}\ \bibnamefont
  {Groot}},\ }\href {https://doi.org/10.1063/1.3276799} {\bibfield  {journal}
  {\bibinfo  {journal} {J. Chem. Phys.}\ }\textbf {\bibinfo {volume} {131}},\
  \bibinfo {pages} {244104} (\bibinfo {year} {2009})}\BibitemShut {NoStop}%
\bibitem [{\citenamefont {Srivastava}\ \emph {et~al.}(2021)\citenamefont
  {Srivastava}, \citenamefont {Roberts}, \citenamefont {Clemmer}, \citenamefont
  {Silbert}, \citenamefont {Lechman},\ and\ \citenamefont
  {Grest}}]{srivastava2021}%
  \BibitemOpen
  \bibfield  {author} {\bibinfo {author} {\bibfnamefont {I.}~\bibnamefont
  {Srivastava}}, \bibinfo {author} {\bibfnamefont {S.~A.}\ \bibnamefont
  {Roberts}}, \bibinfo {author} {\bibfnamefont {J.~T.}\ \bibnamefont
  {Clemmer}}, \bibinfo {author} {\bibfnamefont {L.~E.}\ \bibnamefont
  {Silbert}}, \bibinfo {author} {\bibfnamefont {J.~B.}\ \bibnamefont
  {Lechman}},\ and\ \bibinfo {author} {\bibfnamefont {G.~S.}\ \bibnamefont
  {Grest}},\ }\href {https://doi.org/10.1103/PhysRevResearch.3.L032042}
  {\bibfield  {journal} {\bibinfo  {journal} {Phys. Rev. Res.}\ }\textbf
  {\bibinfo {volume} {3}},\ \bibinfo {pages} {L032042} (\bibinfo {year}
  {2021})}\BibitemShut {NoStop}%
\bibitem [{\citenamefont {Danisch}\ \emph {et~al.}(2010)\citenamefont
  {Danisch}, \citenamefont {Jin},\ and\ \citenamefont {Makse}}]{danisch2010}%
  \BibitemOpen
  \bibfield  {author} {\bibinfo {author} {\bibfnamefont {M.}~\bibnamefont
  {Danisch}}, \bibinfo {author} {\bibfnamefont {Y.}~\bibnamefont {Jin}},\ and\
  \bibinfo {author} {\bibfnamefont {H.~A.}\ \bibnamefont {Makse}},\ }\href
  {https://doi.org/10.1103/PhysRevE.81.051303} {\bibfield  {journal} {\bibinfo
  {journal} {Phys. Rev. E}\ }\textbf {\bibinfo {volume} {81}},\ \bibinfo
  {pages} {051303} (\bibinfo {year} {2010})}\BibitemShut {NoStop}%
\bibitem [{\citenamefont {Hermes}\ and\ \citenamefont
  {Dijkstra}(2010)}]{hermes2010}%
  \BibitemOpen
  \bibfield  {author} {\bibinfo {author} {\bibfnamefont {M.}~\bibnamefont
  {Hermes}}\ and\ \bibinfo {author} {\bibfnamefont {M.}~\bibnamefont
  {Dijkstra}},\ }\href {https://doi.org/10.1209/0295-5075/89/38005} {\bibfield
  {journal} {\bibinfo  {journal} {Europhys. Lett.}\ }\textbf {\bibinfo {volume}
  {89}},\ \bibinfo {pages} {38005} (\bibinfo {year} {2010})}\BibitemShut
  {NoStop}%
\bibitem [{\citenamefont {Desmond}\ and\ \citenamefont
  {Weeks}(2014)}]{desmond2014}%
  \BibitemOpen
  \bibfield  {author} {\bibinfo {author} {\bibfnamefont {K.~W.}\ \bibnamefont
  {Desmond}}\ and\ \bibinfo {author} {\bibfnamefont {E.~R.}\ \bibnamefont
  {Weeks}},\ }\href {https://doi.org/10.1103/PhysRevE.90.022204} {\bibfield
  {journal} {\bibinfo  {journal} {Phys. Rev. E}\ }\textbf {\bibinfo {volume}
  {90}},\ \bibinfo {pages} {022204} (\bibinfo {year} {2014})}\BibitemShut
  {NoStop}%
\bibitem [{\citenamefont {Baranau}\ and\ \citenamefont
  {Tallarek}(2014)}]{baranau2014}%
  \BibitemOpen
  \bibfield  {author} {\bibinfo {author} {\bibfnamefont {V.}~\bibnamefont
  {Baranau}}\ and\ \bibinfo {author} {\bibfnamefont {U.}~\bibnamefont
  {Tallarek}},\ }\href {https://doi.org/10.1039/c3sm52959b} {\bibfield
  {journal} {\bibinfo  {journal} {Soft Matter}\ }\textbf {\bibinfo {volume}
  {10}},\ \bibinfo {pages} {3826} (\bibinfo {year} {2014})}\BibitemShut
  {NoStop}%
\bibitem [{\citenamefont {Cantor}\ \emph {et~al.}(2018)\citenamefont {Cantor},
  \citenamefont {Az{\'{e}}ma}, \citenamefont {Sornay},\ and\ \citenamefont
  {Radjai}}]{cantor2018}%
  \BibitemOpen
  \bibfield  {author} {\bibinfo {author} {\bibfnamefont {D.}~\bibnamefont
  {Cantor}}, \bibinfo {author} {\bibfnamefont {E.}~\bibnamefont {Az{\'{e}}ma}},
  \bibinfo {author} {\bibfnamefont {P.}~\bibnamefont {Sornay}},\ and\ \bibinfo
  {author} {\bibfnamefont {F.}~\bibnamefont {Radjai}},\ }\href
  {https://doi.org/10.1103/PhysRevE.98.052910} {\bibfield  {journal} {\bibinfo
  {journal} {Phys. Rev. E}\ }\textbf {\bibinfo {volume} {98}},\ \bibinfo
  {pages} {052910} (\bibinfo {year} {2018})}\BibitemShut {NoStop}%
\bibitem [{\citenamefont {Mutabaruka}\ \emph {et~al.}(2019)\citenamefont
  {Mutabaruka}, \citenamefont {Taiebat}, \citenamefont {Pellenq},\ and\
  \citenamefont {Radjai}}]{mutabaruka2019}%
  \BibitemOpen
  \bibfield  {author} {\bibinfo {author} {\bibfnamefont {P.}~\bibnamefont
  {Mutabaruka}}, \bibinfo {author} {\bibfnamefont {M.}~\bibnamefont {Taiebat}},
  \bibinfo {author} {\bibfnamefont {R.~J.-M.}\ \bibnamefont {Pellenq}},\ and\
  \bibinfo {author} {\bibfnamefont {F.}~\bibnamefont {Radjai}},\ }\href
  {https://doi.org/10.1103/PhysRevE.100.042906} {\bibfield  {journal} {\bibinfo
   {journal} {Phys. Rev. E}\ }\textbf {\bibinfo {volume} {100}},\ \bibinfo
  {pages} {042906} (\bibinfo {year} {2019})}\BibitemShut {NoStop}%
\bibitem [{\citenamefont {Oquendo-Pati{\~{n}}o}\ and\ \citenamefont
  {Estrada}(2020)}]{oquendo2020}%
  \BibitemOpen
  \bibfield  {author} {\bibinfo {author} {\bibfnamefont {W.~F.}\ \bibnamefont
  {Oquendo-Pati{\~{n}}o}}\ and\ \bibinfo {author} {\bibfnamefont
  {N.}~\bibnamefont {Estrada}},\ }\href
  {https://doi.org/10.1007/s10035-020-01043-9} {\bibfield  {journal} {\bibinfo
  {journal} {Granular Matter}\ }\textbf {\bibinfo {volume} {22}},\ \bibinfo
  {pages} {75} (\bibinfo {year} {2020})}\BibitemShut {NoStop}%
\bibitem [{\citenamefont {Oquendo-Pati{\~{n}}o}\ and\ \citenamefont
  {Estrada}(2021)}]{oquendo2021}%
  \BibitemOpen
  \bibfield  {author} {\bibinfo {author} {\bibfnamefont {W.~F.}\ \bibnamefont
  {Oquendo-Pati{\~{n}}o}}\ and\ \bibinfo {author} {\bibfnamefont
  {N.}~\bibnamefont {Estrada}},\ }\href
  {https://doi.org/10.1051/epjconf/202124902003} {\bibfield  {journal}
  {\bibinfo  {journal} {EPJ Web Conf.}\ }\textbf {\bibinfo {volume} {249}},\
  \bibinfo {pages} {02003} (\bibinfo {year} {2021})}\BibitemShut {NoStop}%
\bibitem [{\citenamefont {Oquendo-Pati{\~{n}}o}\ and\ \citenamefont
  {Estrada}(2022)}]{oquendo2022}%
  \BibitemOpen
  \bibfield  {author} {\bibinfo {author} {\bibfnamefont {W.~F.}\ \bibnamefont
  {Oquendo-Pati{\~{n}}o}}\ and\ \bibinfo {author} {\bibfnamefont
  {N.}~\bibnamefont {Estrada}},\ }\href
  {https://doi.org/10.1103/physreve.105.064901} {\bibfield  {journal} {\bibinfo
   {journal} {Phys. Rev. E}\ }\textbf {\bibinfo {volume} {105}},\ \bibinfo
  {pages} {1} (\bibinfo {year} {2022})}\BibitemShut {NoStop}%
\bibitem [{\citenamefont {Herman}(2013)}]{herman2013}%
  \BibitemOpen
  \bibfield  {author} {\bibinfo {author} {\bibfnamefont {A.}~\bibnamefont
  {Herman}},\ }\href {https://doi.org/10.3390/e15114802} {\bibfield  {journal}
  {\bibinfo  {journal} {Entropy}\ }\textbf {\bibinfo {volume} {15}},\ \bibinfo
  {pages} {4802} (\bibinfo {year} {2013})}\BibitemShut {NoStop}%
\bibitem [{\citenamefont {Filgueira}\ \emph {et~al.}(2006)\citenamefont
  {Filgueira}, \citenamefont {Fournier}, \citenamefont {Cerisola},
  \citenamefont {Gelati},\ and\ \citenamefont {Garc{\'{i}}a}}]{filgueira2006}%
  \BibitemOpen
  \bibfield  {author} {\bibinfo {author} {\bibfnamefont {R.~R.}\ \bibnamefont
  {Filgueira}}, \bibinfo {author} {\bibfnamefont {L.~L.}\ \bibnamefont
  {Fournier}}, \bibinfo {author} {\bibfnamefont {C.~I.}\ \bibnamefont
  {Cerisola}}, \bibinfo {author} {\bibfnamefont {P.}~\bibnamefont {Gelati}},\
  and\ \bibinfo {author} {\bibfnamefont {M.~G.}\ \bibnamefont {Garc{\'{i}}a}},\
  }\href {https://doi.org/10.1016/j.geoderma.2006.03.008} {\bibfield  {journal}
  {\bibinfo  {journal} {Geoderma}\ }\textbf {\bibinfo {volume} {134}},\
  \bibinfo {pages} {327} (\bibinfo {year} {2006})}\BibitemShut {NoStop}%
\bibitem [{\citenamefont {Ben-Nun}\ and\ \citenamefont
  {Einav}(2010)}]{ben-nun2010}%
  \BibitemOpen
  \bibfield  {author} {\bibinfo {author} {\bibfnamefont {O.}~\bibnamefont
  {Ben-Nun}}\ and\ \bibinfo {author} {\bibfnamefont {I.}~\bibnamefont
  {Einav}},\ }\href {https://doi.org/10.1098/rsta.2009.0205} {\bibfield
  {journal} {\bibinfo  {journal} {Philos. Trans. R. Soc., A}\ }\textbf
  {\bibinfo {volume} {368}},\ \bibinfo {pages} {231} (\bibinfo {year}
  {2010})}\BibitemShut {NoStop}%
\bibitem [{\citenamefont {Minh}\ and\ \citenamefont {Cheng}(2013)}]{minh2013}%
  \BibitemOpen
  \bibfield  {author} {\bibinfo {author} {\bibfnamefont {N.~H.}\ \bibnamefont
  {Minh}}\ and\ \bibinfo {author} {\bibfnamefont {Y.~P.}\ \bibnamefont
  {Cheng}},\ }\href {https://doi.org/10.1680/geot.10.P.058} {\bibfield
  {journal} {\bibinfo  {journal} {Geotechnique}\ }\textbf {\bibinfo {volume}
  {63}},\ \bibinfo {pages} {44} (\bibinfo {year} {2013})}\BibitemShut {NoStop}%
\bibitem [{\citenamefont {de~Bono}\ and\ \citenamefont
  {McDowell}(2020)}]{debono2020}%
  \BibitemOpen
  \bibfield  {author} {\bibinfo {author} {\bibfnamefont {J.~P.}\ \bibnamefont
  {de~Bono}}\ and\ \bibinfo {author} {\bibfnamefont {G.~R.}\ \bibnamefont
  {McDowell}},\ }\href {https://doi.org/10.1016/j.ijsolstr.2018.07.011}
  {\bibfield  {journal} {\bibinfo  {journal} {Int. J. Solids Struct.}\ }\textbf
  {\bibinfo {volume} {187}},\ \bibinfo {pages} {133} (\bibinfo {year}
  {2020})}\BibitemShut {NoStop}%
\bibitem [{\citenamefont {Borkovec}\ \emph {et~al.}(1994)\citenamefont
  {Borkovec}, \citenamefont {{De Paris}},\ and\ \citenamefont
  {Peikert}}]{borkovec1994}%
  \BibitemOpen
  \bibfield  {author} {\bibinfo {author} {\bibfnamefont {M.}~\bibnamefont
  {Borkovec}}, \bibinfo {author} {\bibfnamefont {W.}~\bibnamefont {{De
  Paris}}},\ and\ \bibinfo {author} {\bibfnamefont {R.}~\bibnamefont
  {Peikert}},\ }\href {https://doi.org/10.1142/S0218348X94000739} {\bibfield
  {journal} {\bibinfo  {journal} {Fractals}\ }\textbf {\bibinfo {volume}
  {02}},\ \bibinfo {pages} {521} (\bibinfo {year} {1994})}\BibitemShut
  {NoStop}%
\bibitem [{\citenamefont {Anishchik}\ and\ \citenamefont
  {Medvedev}(1995)}]{anishchik1995}%
  \BibitemOpen
  \bibfield  {author} {\bibinfo {author} {\bibfnamefont {S.~V.}\ \bibnamefont
  {Anishchik}}\ and\ \bibinfo {author} {\bibfnamefont {N.~N.}\ \bibnamefont
  {Medvedev}},\ }\href {https://doi.org/10.1103/PhysRevLett.75.4314} {\bibfield
   {journal} {\bibinfo  {journal} {Phys. Rev. Lett.}\ }\textbf {\bibinfo
  {volume} {75}},\ \bibinfo {pages} {4314} (\bibinfo {year}
  {1995})}\BibitemShut {NoStop}%
\bibitem [{\citenamefont {Herrmann}\ \emph {et~al.}(2003)\citenamefont
  {Herrmann}, \citenamefont {{Mahmoodi Baram}},\ and\ \citenamefont
  {Wackenhut}}]{herrmann2003}%
  \BibitemOpen
  \bibfield  {author} {\bibinfo {author} {\bibfnamefont {H.}~\bibnamefont
  {Herrmann}}, \bibinfo {author} {\bibfnamefont {R.}~\bibnamefont {{Mahmoodi
  Baram}}},\ and\ \bibinfo {author} {\bibfnamefont {M.}~\bibnamefont
  {Wackenhut}},\ }\href {https://doi.org/10.1016/j.physa.2003.08.023}
  {\bibfield  {journal} {\bibinfo  {journal} {Phys. A}\ }\textbf {\bibinfo
  {volume} {330}},\ \bibinfo {pages} {77} (\bibinfo {year} {2003})}\BibitemShut
  {NoStop}%
\bibitem [{\citenamefont {Varrato}\ and\ \citenamefont
  {Foffi}(2011)}]{varrato2011}%
  \BibitemOpen
  \bibfield  {author} {\bibinfo {author} {\bibfnamefont {F.}~\bibnamefont
  {Varrato}}\ and\ \bibinfo {author} {\bibfnamefont {G.}~\bibnamefont
  {Foffi}},\ }\href {https://doi.org/10.1080/00268976.2011.640039} {\bibfield
  {journal} {\bibinfo  {journal} {Mol. Phys.}\ }\textbf {\bibinfo {volume}
  {109}},\ \bibinfo {pages} {2923} (\bibinfo {year} {2011})}\BibitemShut
  {NoStop}%
\bibitem [{\citenamefont {Plimpton}(1995)}]{plimpton1995}%
  \BibitemOpen
  \bibfield  {author} {\bibinfo {author} {\bibfnamefont {S.}~\bibnamefont
  {Plimpton}},\ }\href {https://doi.org/https://doi.org/10.1006/jcph.1995.1039}
  {\bibfield  {journal} {\bibinfo  {journal} {J. Comput. Phys.}\ }\textbf
  {\bibinfo {volume} {117}},\ \bibinfo {pages} {1} (\bibinfo {year}
  {1995})}\BibitemShut {NoStop}%
\bibitem [{\citenamefont {Thompson}\ \emph {et~al.}(2022)\citenamefont
  {Thompson}, \citenamefont {Aktulga}, \citenamefont {Berger}, \citenamefont
  {Bolintineanu}, \citenamefont {Brown}, \citenamefont {Crozier}, \citenamefont
  {{in 't Veld}}, \citenamefont {Kohlmeyer}, \citenamefont {Moore},
  \citenamefont {Nguyen}, \citenamefont {Shan}, \citenamefont {Stevens},
  \citenamefont {Tranchida}, \citenamefont {Trott},\ and\ \citenamefont
  {Plimpton}}]{thompson2022}%
  \BibitemOpen
  \bibfield  {author} {\bibinfo {author} {\bibfnamefont {A.~P.}\ \bibnamefont
  {Thompson}}, \bibinfo {author} {\bibfnamefont {H.~M.}\ \bibnamefont
  {Aktulga}}, \bibinfo {author} {\bibfnamefont {R.}~\bibnamefont {Berger}},
  \bibinfo {author} {\bibfnamefont {D.~S.}\ \bibnamefont {Bolintineanu}},
  \bibinfo {author} {\bibfnamefont {W.~M.}\ \bibnamefont {Brown}}, \bibinfo
  {author} {\bibfnamefont {P.~S.}\ \bibnamefont {Crozier}}, \bibinfo {author}
  {\bibfnamefont {P.~J.}\ \bibnamefont {{in 't Veld}}}, \bibinfo {author}
  {\bibfnamefont {A.}~\bibnamefont {Kohlmeyer}}, \bibinfo {author}
  {\bibfnamefont {S.~G.}\ \bibnamefont {Moore}}, \bibinfo {author}
  {\bibfnamefont {T.~D.}\ \bibnamefont {Nguyen}}, \bibinfo {author}
  {\bibfnamefont {R.}~\bibnamefont {Shan}}, \bibinfo {author} {\bibfnamefont
  {M.~J.}\ \bibnamefont {Stevens}}, \bibinfo {author} {\bibfnamefont
  {J.}~\bibnamefont {Tranchida}}, \bibinfo {author} {\bibfnamefont
  {C.}~\bibnamefont {Trott}},\ and\ \bibinfo {author} {\bibfnamefont {S.~J.}\
  \bibnamefont {Plimpton}},\ }\href {https://doi.org/10.1016/j.cpc.2021.108171}
  {\bibfield  {journal} {\bibinfo  {journal} {Comput. Phys. Commun.}\ }\textbf
  {\bibinfo {volume} {271}},\ \bibinfo {pages} {108171} (\bibinfo {year}
  {2022})}\BibitemShut {NoStop}%
\bibitem [{\citenamefont {{in 't Veld}}\ \emph {et~al.}(2008)\citenamefont {{in
  't Veld}}, \citenamefont {Plimpton},\ and\ \citenamefont {Grest}}]{veld2008}%
  \BibitemOpen
  \bibfield  {author} {\bibinfo {author} {\bibfnamefont {P.~J.}\ \bibnamefont
  {{in 't Veld}}}, \bibinfo {author} {\bibfnamefont {S.~J.}\ \bibnamefont
  {Plimpton}},\ and\ \bibinfo {author} {\bibfnamefont {G.~S.}\ \bibnamefont
  {Grest}},\ }\href {https://doi.org/https://doi.org/10.1016/j.cpc.2008.03.005}
  {\bibfield  {journal} {\bibinfo  {journal} {Comput. Phys. Commun.}\ }\textbf
  {\bibinfo {volume} {179}},\ \bibinfo {pages} {320} (\bibinfo {year}
  {2008})}\BibitemShut {NoStop}%
\bibitem [{\citenamefont {Ogarko}\ and\ \citenamefont
  {Luding}(2012)}]{ogarko2012}%
  \BibitemOpen
  \bibfield  {author} {\bibinfo {author} {\bibfnamefont {V.}~\bibnamefont
  {Ogarko}}\ and\ \bibinfo {author} {\bibfnamefont {S.}~\bibnamefont
  {Luding}},\ }\href
  {https://doi.org/https://doi.org/10.1016/j.cpc.2011.12.019} {\bibfield
  {journal} {\bibinfo  {journal} {Comput. Phys. Commun.}\ }\textbf {\bibinfo
  {volume} {183}},\ \bibinfo {pages} {931} (\bibinfo {year}
  {2012})}\BibitemShut {NoStop}%
\bibitem [{\citenamefont {Shire}\ \emph {et~al.}(2021)\citenamefont {Shire},
  \citenamefont {Hanley},\ and\ \citenamefont {Stratford}}]{shire2021}%
  \BibitemOpen
  \bibfield  {author} {\bibinfo {author} {\bibfnamefont {T.}~\bibnamefont
  {Shire}}, \bibinfo {author} {\bibfnamefont {K.~J.}\ \bibnamefont {Hanley}},\
  and\ \bibinfo {author} {\bibfnamefont {K.}~\bibnamefont {Stratford}},\ }\href
  {https://doi.org/10.1007/s40571-020-00361-2} {\bibfield  {journal} {\bibinfo
  {journal} {Comp. Part. Mech.}\ }\textbf {\bibinfo {volume} {8}},\ \bibinfo
  {pages} {653} (\bibinfo {year} {2021})}\BibitemShut {NoStop}%
\bibitem [{\citenamefont {Dagois-Bohy}\ \emph {et~al.}(2012)\citenamefont
  {Dagois-Bohy}, \citenamefont {Tighe}, \citenamefont {Simon}, \citenamefont
  {Henkes},\ and\ \citenamefont {van Hecke}}]{dagois2012}%
  \BibitemOpen
  \bibfield  {author} {\bibinfo {author} {\bibfnamefont {S.}~\bibnamefont
  {Dagois-Bohy}}, \bibinfo {author} {\bibfnamefont {B.~P.}\ \bibnamefont
  {Tighe}}, \bibinfo {author} {\bibfnamefont {J.}~\bibnamefont {Simon}},
  \bibinfo {author} {\bibfnamefont {S.}~\bibnamefont {Henkes}},\ and\ \bibinfo
  {author} {\bibfnamefont {M.}~\bibnamefont {van Hecke}},\ }\href
  {https://doi.org/10.1103/PhysRevLett.109.095703} {\bibfield  {journal}
  {\bibinfo  {journal} {Phys. Rev. Lett.}\ }\textbf {\bibinfo {volume} {109}},\
  \bibinfo {pages} {095703} (\bibinfo {year} {2012})}\BibitemShut {NoStop}%
\bibitem [{\citenamefont {Smith}\ \emph {et~al.}(2014)\citenamefont {Smith},
  \citenamefont {Srivastava}, \citenamefont {Fisher},\ and\ \citenamefont
  {Alam}}]{smith2014}%
  \BibitemOpen
  \bibfield  {author} {\bibinfo {author} {\bibfnamefont {K.~C.}\ \bibnamefont
  {Smith}}, \bibinfo {author} {\bibfnamefont {I.}~\bibnamefont {Srivastava}},
  \bibinfo {author} {\bibfnamefont {T.~S.}\ \bibnamefont {Fisher}},\ and\
  \bibinfo {author} {\bibfnamefont {M.}~\bibnamefont {Alam}},\ }\href
  {https://doi.org/10.1103/PhysRevE.89.042203} {\bibfield  {journal} {\bibinfo
  {journal} {Phys. Rev. E}\ }\textbf {\bibinfo {volume} {89}},\ \bibinfo
  {pages} {042203} (\bibinfo {year} {2014})}\BibitemShut {NoStop}%
\bibitem [{\citenamefont {Santos}\ \emph {et~al.}(2020)\citenamefont {Santos},
  \citenamefont {Bolintineanu}, \citenamefont {Grest}, \citenamefont {Lechman},
  \citenamefont {Plimpton}, \citenamefont {Srivastava},\ and\ \citenamefont
  {Silbert}}]{santos2020}%
  \BibitemOpen
  \bibfield  {author} {\bibinfo {author} {\bibfnamefont {A.~P.}\ \bibnamefont
  {Santos}}, \bibinfo {author} {\bibfnamefont {D.~S.}\ \bibnamefont
  {Bolintineanu}}, \bibinfo {author} {\bibfnamefont {G.~S.}\ \bibnamefont
  {Grest}}, \bibinfo {author} {\bibfnamefont {J.~B.}\ \bibnamefont {Lechman}},
  \bibinfo {author} {\bibfnamefont {S.~J.}\ \bibnamefont {Plimpton}}, \bibinfo
  {author} {\bibfnamefont {I.}~\bibnamefont {Srivastava}},\ and\ \bibinfo
  {author} {\bibfnamefont {L.~E.}\ \bibnamefont {Silbert}},\ }\href
  {https://doi.org/10.1103/PhysRevE.102.032903} {\bibfield  {journal} {\bibinfo
   {journal} {Phys. Rev. E}\ }\textbf {\bibinfo {volume} {102}},\ \bibinfo
  {pages} {032903} (\bibinfo {year} {2020})}\BibitemShut {NoStop}%
\bibitem [{\citenamefont {Silbert}\ \emph {et~al.}(2001)\citenamefont
  {Silbert}, \citenamefont {Erta\c{s}}, \citenamefont {Grest}, \citenamefont
  {Halsey}, \citenamefont {Levine},\ and\ \citenamefont
  {Plimpton}}]{silbert2001}%
  \BibitemOpen
  \bibfield  {author} {\bibinfo {author} {\bibfnamefont {L.~E.}\ \bibnamefont
  {Silbert}}, \bibinfo {author} {\bibfnamefont {D.}~\bibnamefont {Erta\c{s}}},
  \bibinfo {author} {\bibfnamefont {G.~S.}\ \bibnamefont {Grest}}, \bibinfo
  {author} {\bibfnamefont {T.~C.}\ \bibnamefont {Halsey}}, \bibinfo {author}
  {\bibfnamefont {D.}~\bibnamefont {Levine}},\ and\ \bibinfo {author}
  {\bibfnamefont {S.~J.}\ \bibnamefont {Plimpton}},\ }\href
  {https://doi.org/10.1103/PhysRevE.64.051302} {\bibfield  {journal} {\bibinfo
  {journal} {Phys. Rev. E}\ }\textbf {\bibinfo {volume} {64}},\ \bibinfo
  {pages} {051302} (\bibinfo {year} {2001})}\BibitemShut {NoStop}%
\bibitem [{\citenamefont {Stukowski}(2009)}]{ovito}%
  \BibitemOpen
  \bibfield  {author} {\bibinfo {author} {\bibfnamefont {A.}~\bibnamefont
  {Stukowski}},\ }\href {https://doi.org/10.1088/0965-0393/18/1/015012}
  {\bibfield  {journal} {\bibinfo  {journal} {Modell. Simul. Mater. Sci. Eng.}\
  }\textbf {\bibinfo {volume} {18}},\ \bibinfo {pages} {015012} (\bibinfo
  {year} {2009})}\BibitemShut {NoStop}%
\bibitem [{\citenamefont {Stratford}\ \emph {et~al.}(2018)\citenamefont
  {Stratford}, \citenamefont {Shire},\ and\ \citenamefont
  {Hanley}}]{stratford2018}%
  \BibitemOpen
  \bibfield  {author} {\bibinfo {author} {\bibfnamefont {K.}~\bibnamefont
  {Stratford}}, \bibinfo {author} {\bibfnamefont {T.}~\bibnamefont {Shire}},\
  and\ \bibinfo {author} {\bibfnamefont {K.}~\bibnamefont {Hanley}},\
  }\href@noop {} {\emph {\bibinfo {title} {Implementation of multi-level
  contact detection in {LAMMPS}}}},\ \bibinfo {type} {Tech. Rep.}\ \bibinfo
  {number} {eCSE12-09}\ (\bibinfo  {institution} {University of Edinburgh
  (United Kingdom)},\ \bibinfo {year} {2018})\BibitemShut {NoStop}%
\bibitem [{Note1()}]{Note1}%
  \BibitemOpen
  \bibinfo {note} {These capabilities are described in the documentation found
  at https://docs.lammps.org/neighbor.html. An example input script in.powerlaw
  is included with the LAMMPS distribution in the examples/multi subdirectory.
  It is a demo of shearing a 2D packing of particles with power-law-distributed
  sizes using neighbor list options similar to those leveraged in this
  work.}\BibitemShut {Stop}%
\bibitem [{\citenamefont {Liu}\ and\ \citenamefont {Nagel}(2010)}]{liu2010}%
  \BibitemOpen
  \bibfield  {author} {\bibinfo {author} {\bibfnamefont {A.~J.}\ \bibnamefont
  {Liu}}\ and\ \bibinfo {author} {\bibfnamefont {S.~R.}\ \bibnamefont
  {Nagel}},\ }\href {https://doi.org/10.1146/annurev-conmatphys-070909-104045}
  {\bibfield  {journal} {\bibinfo  {journal} {Annu. Rev. Condens. Matter
  Phys.}\ }\textbf {\bibinfo {volume} {1}},\ \bibinfo {pages} {347} (\bibinfo
  {year} {2010})}\BibitemShut {NoStop}%
\bibitem [{\citenamefont {Prasad}\ \emph {et~al.}(2017)\citenamefont {Prasad},
  \citenamefont {Santangelo},\ and\ \citenamefont {Grason}}]{prasad2017}%
  \BibitemOpen
  \bibfield  {author} {\bibinfo {author} {\bibfnamefont {I.}~\bibnamefont
  {Prasad}}, \bibinfo {author} {\bibfnamefont {C.}~\bibnamefont {Santangelo}},\
  and\ \bibinfo {author} {\bibfnamefont {G.}~\bibnamefont {Grason}},\ }\href
  {https://doi.org/10.1103/PhysRevE.96.052905} {\bibfield  {journal} {\bibinfo
  {journal} {Phys. Rev. E}\ }\textbf {\bibinfo {volume} {96}},\ \bibinfo
  {pages} {052905} (\bibinfo {year} {2017})}\BibitemShut {NoStop}%
\bibitem [{\citenamefont {Aste}(1996)}]{aste1996}%
  \BibitemOpen
  \bibfield  {author} {\bibinfo {author} {\bibfnamefont {T.}~\bibnamefont
  {Aste}},\ }\href {https://doi.org/10.1103/PhysRevE.53.2571} {\bibfield
  {journal} {\bibinfo  {journal} {Phys. Rev. E}\ }\textbf {\bibinfo {volume}
  {53}},\ \bibinfo {pages} {2571} (\bibinfo {year} {1996})}\BibitemShut
  {NoStop}%
\bibitem [{\citenamefont {Roux}(2000)}]{roux2000}%
  \BibitemOpen
  \bibfield  {author} {\bibinfo {author} {\bibfnamefont {J.-N.}\ \bibnamefont
  {Roux}},\ }\href {https://doi.org/10.1103/PhysRevE.61.6802} {\bibfield
  {journal} {\bibinfo  {journal} {Phys. Rev. E}\ }\textbf {\bibinfo {volume}
  {61}},\ \bibinfo {pages} {6802} (\bibinfo {year} {2000})}\BibitemShut
  {NoStop}%
\bibitem [{\citenamefont {Donev}\ \emph {et~al.}(2004)\citenamefont {Donev},
  \citenamefont {Torquato}, \citenamefont {Stillinger},\ and\ \citenamefont
  {Connelly}}]{donev2004}%
  \BibitemOpen
  \bibfield  {author} {\bibinfo {author} {\bibfnamefont {A.}~\bibnamefont
  {Donev}}, \bibinfo {author} {\bibfnamefont {S.}~\bibnamefont {Torquato}},
  \bibinfo {author} {\bibfnamefont {F.~H.}\ \bibnamefont {Stillinger}},\ and\
  \bibinfo {author} {\bibfnamefont {R.}~\bibnamefont {Connelly}},\ }\href
  {https://doi.org/10.1016/j.jcp.2003.11.022} {\bibfield  {journal} {\bibinfo
  {journal} {J. Comput. Phys.}\ }\textbf {\bibinfo {volume} {197}},\ \bibinfo
  {pages} {139} (\bibinfo {year} {2004})}\BibitemShut {NoStop}%
\bibitem [{\citenamefont {Corwin}\ \emph {et~al.}(2010)\citenamefont {Corwin},
  \citenamefont {Clusel}, \citenamefont {Siemens},\ and\ \citenamefont
  {Bruji{\'{c}}}}]{corwin2010}%
  \BibitemOpen
  \bibfield  {author} {\bibinfo {author} {\bibfnamefont {E.~I.}\ \bibnamefont
  {Corwin}}, \bibinfo {author} {\bibfnamefont {M.}~\bibnamefont {Clusel}},
  \bibinfo {author} {\bibfnamefont {A.~O.~N.}\ \bibnamefont {Siemens}},\ and\
  \bibinfo {author} {\bibfnamefont {J.}~\bibnamefont {Bruji{\'{c}}}},\ }\href
  {https://doi.org/10.1039/c000984a} {\bibfield  {journal} {\bibinfo  {journal}
  {Soft Matter}\ }\textbf {\bibinfo {volume} {6}},\ \bibinfo {pages} {2949}
  (\bibinfo {year} {2010})}\BibitemShut {NoStop}%
\bibitem [{\citenamefont {Otsubo}(2016)}]{otsubo2016}%
  \BibitemOpen
  \bibfield  {author} {\bibinfo {author} {\bibfnamefont {M.}~\bibnamefont
  {Otsubo}},\ }\emph {\bibinfo {title} {Particle scale analysis of soil
  stiffness and elastic wave propagation}},\ \href@noop {} {Ph.D. thesis},\
  \bibinfo  {school} {Imperial College London} (\bibinfo {year}
  {2016})\BibitemShut {NoStop}%
\bibitem [{\citenamefont {Liu}\ \emph {et~al.}(2021)\citenamefont {Liu},
  \citenamefont {O'Sullivan},\ and\ \citenamefont {Carraro}}]{liu2021}%
  \BibitemOpen
  \bibfield  {author} {\bibinfo {author} {\bibfnamefont {D.}~\bibnamefont
  {Liu}}, \bibinfo {author} {\bibfnamefont {C.}~\bibnamefont {O'Sullivan}},\
  and\ \bibinfo {author} {\bibfnamefont {J.~A.~H.}\ \bibnamefont {Carraro}},\
  }\href {https://doi.org/10.1061/(ASCE)GT.1943-5606.0002466} {\bibfield
  {journal} {\bibinfo  {journal} {J. Geotech. Geoenviron. Eng.}\ }\textbf
  {\bibinfo {volume} {147}},\ \bibinfo {pages} {04020182} (\bibinfo {year}
  {2021})}\BibitemShut {NoStop}%
\bibitem [{\citenamefont {Krijgsman}\ \emph {et~al.}(2014)\citenamefont
  {Krijgsman}, \citenamefont {Ogarko},\ and\ \citenamefont
  {Luding}}]{krijgsman2014}%
  \BibitemOpen
  \bibfield  {author} {\bibinfo {author} {\bibfnamefont {D.}~\bibnamefont
  {Krijgsman}}, \bibinfo {author} {\bibfnamefont {V.}~\bibnamefont {Ogarko}},\
  and\ \bibinfo {author} {\bibfnamefont {S.}~\bibnamefont {Luding}},\ }\href
  {https://doi.org/10.1007/s40571-014-0020-9} {\bibfield  {journal} {\bibinfo
  {journal} {Comp. Part. Mech.}\ }\textbf {\bibinfo {volume} {1}},\ \bibinfo
  {pages} {357} (\bibinfo {year} {2014})}\BibitemShut {NoStop}%
\bibitem [{Note2()}]{Note2}%
  \BibitemOpen
  \bibinfo {note} {Some aspects of neighbor list construction in LAMMPS, such
  as the addition of a skin to the interaction distance and the existence of
  full and half stencils, are not described in this Appendix as they do not
  affect the underlying scaling of costs with $\lambda $.}\BibitemShut {Stop}%
\end{thebibliography}%
\end{document}